\documentclass[12pt]{amsart}
\usepackage{amssymb,amsfonts,amsmath}
\usepackage{graphicx}
\usepackage{epsfig}
\usepackage{float}

\newtheorem{theorem}{Theorem}[section]
\newtheorem{lemma}{Lemma}[section]
\newtheorem{definition}{Definition}[section]

\numberwithin{equation}{section} \numberwithin{theorem}{section}

\newcounter{labelflag} 

\newcommand{\Label}[1]{
                       \ifnum\thelabelflag=1
                          \ifmmode
                             \makebox[0in][l]{\qquad\fbox{\rm#1}}
                          \else
                             \marginpar{\vspace{0.7\baselineskip}
                                        \hspace{-1.1\textwidth}
                                        \fbox{\rm#1}}
                          \fi
                       \fi
                       \label{#1}
                      }
\newcommand{\be}{\begin{equation}}
\newcommand{\ee}{\end{equation}}
\newcommand{\dom}{\mathrm{dom}}
\renewcommand{\det}{\mathrm{det}}
\newcommand{\tr}{\mathrm{tr}}
\newcommand{\mA}{\mathcal A}
\newcommand{\R}{\mathbf R}
\newcommand{\half}{\textstyle{\frac12}}
\newcommand{\onethird}{\textstyle{\frac13}}
\newcommand{\twothirds}{\textstyle{\frac23}}

\begin{document}
\bibliographystyle{siam}

\title[Formation and Persistence of Patterns]
{Formation and Persistence of Spatiotemporal Turing Patterns}

\author[Hans G.~Kaper]{Hans G.~Kaper}
\address[HGK]{Mathematics and Computer Science Division,
Argonne National Laboratory, Argonne, IL 60439; current address:
Division of Mathematical Sciences, National Science Foundation,
Arlington, VA 22230} \email{kaper@mcs.anl.gov; hkaper@nsf.gov}

\author[Shouhong Wang]{Shouhong Wang}
\address[SW]{Department of Mathematics,
Indiana University, Bloomington, IN 47405}
\email{showang@indiana.edu}

\author[Masoud Yari]{Masoud Yari}
\address[MY]{Department of Mathematics,
Indiana University, Bloomington, IN 47405} \email{myari@indiana.edu}

\begin{abstract}
This article is concerned with the stability and long-time dynamics
of structures arising from a structureless state. The paradigm is
suggested by developmental biology, where morphogenesis is thought
to result from a competition between chemical reactions and spatial
diffusion.  A system of two reaction--diffusion equations for the
concentrations of two morphogens is reduced to a finite system of
ordinary differential equations.  The stability of bifurcated
solutions of this system is analyzed, and the long-time asymptotic
behavior of the bifurcated solutions is established rigorously. The
Schnakenberg and Gierer--Meinhardt equations are discussed as
examples.
\end{abstract}

\maketitle

\section{Introduction\label{s-intro}}
Morphogenesis---that is, the development and formation of tissues
and organs---is one of the main mysteries in living organisms.
How does structure emerge from a structureless state
without the apparent action of an external organizing force?
A major factor seems to be the competition between
chemical reactions and spatial diffusion of substances
called morphogens, which are present in the cells.
The idea goes back to the pioneering work of Turing
in 1952~\cite{tu}, who noted that diffusion in a
mixture of chemically reacting morphogens can
cause instability of a spatially uniform steady state
and lead to the formation of spatial patterns;
see the article by Cross and Hohenberg~\cite{cross-h}
and the recent text by Hoyle~\cite{hoyle}
for a comprehensive overview of the theory
of pattern formation and Turing analysis.

Turing's analysis, which is essentially a linear
eigenvalue analysis, has been a basic tool
in the study of nonlinear reaction--diffusion systems;
in fact, it has provided insight into the behavior of
nonlinear systems as well, since the latter can
often be approximated, at least for brief lengths
of time, by linearized systems.
But as time evolves, the nonlinear structure takes
over, and other tools are needed to study the
long-time behavior.
In certain cases, where the existence of
invariant regions for reaction--diffusion
systems can be established,
the solution of a nonlinear system
remains bounded~\cite{smoller};
however, even in these cases it is not known
rigorously whether the patterns persist
in the long run, even though the idea
is supported by many numerical simulations.

The mathematical literature contains
many instances of weakly nonlinear stability
analyses for reaction--diffusion systems.
An early reference is~\cite{h-o},
where a center-manifold approach is used;
other, more recent references are~\cite{w, z-m}.
Sometimes, special techniques have been applied
to the study of Turing patterns in different regimes.
For example, Ref.~\cite{iww} deals with the stability
of symmetric $N$-peaked steady states
for systems where the inhibitor diffuses
much more rapidly than the activator.
We also mention Refs.~\cite{bms, pmm-2005},
which deal with the Schnakenberg model
on a two-dimensional square domain,
where spatially varying diffusion coefficients
cause the removal of the degeneracy
of the Turing bifurcation.

Weakly nonlinear stability analyses can be
justified rigorously on the basis of
modulation theory and a Ginzburg--Landau approximation;
see, for example, Refs.~\cite{bvhs, eck, schn-1, schn-2, vh}.
Murray's monograph~\cite{murray} gives
applications to biological systems
such as animal coat patterns.

The purpose of the present work is
to investigate the nonlinear stability
and persistence of spatiotemporal patterns
on bounded domains.
The investigation is based on recent results
of Ma and Wang~\cite{mw} on attractor
bifurcation for nonlinear equations.
The attractor bifurcation theorem
(see the Appendix, Section~\ref{ss-abt})
sums up the basic features
of a stability-breaking bifurcation.
Starting from the original partial
differential equation, it identifies and
characterizes the local basins of attraction
based on the multiplicity of the eigenvalues
near a bifurcation point.
Thus, the attractor bifurcation theorem
gives the complete picture,
rather than the caricature given
by the amplitude equations.

A second essential feature
of the present investigation
is a center-manifold reduction
to reduce the partial differential equation
to a finite-dimensional dynamical system.
The reduction requires the computation
of the center-manifold function
and the interaction of the higher-order
eigenfunctions with the eigenspace
belonging to the leading eigenvalues.
Such a reduction is inherently difficult,
and for this reason one usually resorts
to a generic form of the reduced equation
which is somewhat detached from the original.
On the other hand, a center-manifold reduction
offers a practical way to find the structure
of the local attractors of the original
partial differential equation.
These attractors completely describe
the local transitions, and their
basins of attraction define the long-time
dynamics associated with the transitions.
Since these are exactly the features of
interest, we have taken this approach
and focused much of our efforts on
the center-manifold reduction.
As a result, we are able to characterize
the types of transitions in terms of
explicitly computable parameters
which depend only on the domain and
the values of the physical parameters
of the system under consideration.

We prove that spatiotemporal patterns
in reaction--diffusion systems
of the attractor--inhibitor type
can arise as the result of a
supercritical (pitchfork) or
subcritical bifurcation.
The former results in a continuous transition,
the latter in a discontinuous transition.
In the case of diffusion on a (bounded) interval
or on a rectangular (non-square) domain,
we prove that the attractor consists of
two points, each with its basin of attraction
(Theorem~\ref{th-1d}, Fig.~\ref{fig3}).
In the case of diffusion on a (bounded) square,
the phase diagram after bifurcation consists of
eight steady-state solutions and their connecting
heteroclinic orbits
(Theorem~\ref{th-2d}, Fig.~\ref{fig.hetorbit}).
The conditions for the stability of these
bifurcated steady states and the heteroclinic
orbits are explicit;
they can be verified in terms of eigenvalues
and eigenvectors.
In the framework of classical bifurcation theory,
such an explicit characterization is very difficult,
if not impossible, and the existence of heteroclinic
orbits for a partial differential equation is often
hard to prove.

Although the focus in this article is on
activator--inhibitor systems, the analysis
is quite general and applies, for example,
to systems consisting of a self-amplifying
activator and a depleted substrate.

Following is an outline of the paper.
In Section~\ref{s-problem},
we formulate the reaction--diffusion problem
for an activator--inhibitor mixture and
rewrite it as an evolution equation
in a function space.
In Section~\ref{s-exchange},
we study the exchange of stability,
which is crucial for the stability
and bifurcation analysis.
The results of the bifurcation analysis
are summarized for the one-dimensional
case in Section~\ref{s-1d} and the
two-dimensional case in Section~\ref{s-2d}.
In Section~\ref{s-examples},
we illustrate the theoretical results
on two examples, namely the Schnakenberg
equation and the Gierer--Meinhardt equations.
Section~\ref{s-conclusions} summarizes our
conclusions.
Appendix~\ref{s-appendix} contains a brief summary
of the attractor bifurcation theory from Ref.~\cite{mw}
and the reduction method introduced in Ref.~\cite{mw-b}.

\section{Statement of the Problem\label{s-problem}}
Consider a mixture of two chemical species
which simultaneously react and diffuse;
one of the species is an activator,
the other an inhibitor
of the chemical reaction.
Their respective concentrations $U$ and $V$
satisfy a system of coupled nonlinear
reaction--diffusion equations,
\begin{equation}
\begin{split}
   U_t &= f(U,V) + d_1 \Delta U , \\
   V_t &= g(U,V) + d_2 \Delta V ,
\end{split}
\Label{eq1-UV}
\end{equation}
subject to no-flux boundary conditions
and given initial conditions.
The functions $f$ and $g$,
which describe the kinetics
of the chemical reaction,
are generally nonlinear
functions of the arguments.
The diffusion coefficients $d_1$ and $d_2$
are constant and positive.

We assume that the system of Eqs.~(\ref{eq1-UV})
admits a uniform steady-state solution
which is positive throughout the domain.
That is, there exist constants
$\bar{u} > 0$ and $\bar{v} > 0$
such that
\begin{equation}
  f(\bar{u}, \bar{v}) = 0 , \quad g(\bar{u}, \bar{v}) = 0 .
\Label{equil}
\end{equation}
We are interested in
solutions that bifurcate
from this equilibrium solution
and, in particular, in their
long-term dynamics,
under the assumption that
the equilibrium solution~(\ref{equil})
is stable in the absence of diffusion.

For the bifurcation analysis,
it is convenient to rescale time and space
and rewrite the system~(\ref{eq1-UV})
in the form
\begin{equation}
\begin{split}
   U_t &= \gamma f(U,V) + \Delta U , \\
   V_t &= \gamma g(U,V) + d \Delta V ,
\Label{eq2-UV}
\end{split}
\end{equation}
where $\gamma= 1/d_1$ and $d = d_2/d_1$.
Thus,
$\gamma$ is a measure of the ratio
of the characteristic times for diffusion
and chemical reaction,
and $d$ is the ratio of
the diffusion coefficients
of the two species.
The above equations are satisfied
on an open bounded domain,
say $\Omega \subset \R^n$ ($n=1,2$),
while $U$ and $V$ satisfy Neumann (no-flux)
boundary conditions on the boundary $\partial\Omega$
of~$\Omega$.

\subsection{Bifurcation Problem\label{ss-bifurcation}}
Let
\begin{equation}
  U = \bar u + u , \quad V = \bar v + v .
\Label{def-uv}
\end{equation}
Since $\bar{u}$ and $\bar{v}$ satisfy
the identities~(\ref{equil}),
we have
\begin{equation}
\begin{split}
  f(U,V) &= f_u (\bar{u}, \bar{v}) u + f_v (\bar{u}, \bar{v}) v + f_1 (u,v) , \\
  g(U,V) &= g_u (\bar{u}, \bar{v}) u + g_v (\bar{u}, \bar{v}) v + g_1 (u,v) ,
\end{split}
\end{equation}
where $f_1$ and $g_1$ incorporate the
higher-order terms in the Taylor expansions.
The functions $u$ and $v$ satisfy the equations
\begin{equation}
\begin{split}
  u_t
  &=
  \Delta u + \gamma (f_u (\bar{u}, \bar{v}) u + f_v (\bar{u}, \bar{v}) v)
  + \gamma f_1 (u,v) , \\
  v_t
  &=
  d \Delta v + \gamma (g_u (\bar{u}, \bar{v}) u + g_v (\bar{u}, \bar{v}) v)
  + \gamma g_1 (u,v) .
\Label{eq-uv}
\end{split}
\end{equation}
Henceforth we omit the arguments $(\bar{u}, \bar{v})$
and use the abbreviations $f_u$ for $f_u(\bar{u}, \bar{v})$,
et cetera.

Since the variables $U$ and $V$ are associated with
the activator and the inhibitor, respectively,
of the chemical reaction, we have the inequalities
\be
  f_u > 0 , \quad g_v < 0 .
\Label{ineq1-fg}
\ee
The equilibrium solution~(\ref{equil})
is stable in the absence of diffusion,
so we also have the inequalities
\begin{equation}
  f_u g_v - f_v g_u > 0 , \quad
  f_u + g_v < 0 .
\Label{ineq2-fg}
\end{equation}
The first inequality in~(\ref{ineq2-fg}),
together with the inequalities~(\ref{ineq1-fg}),
implies that $f_v g_u < 0$.

The problem as stated has two parameters,
$\gamma$ and $d$.
We represent the ordered pair
by a single symbol,
$\lambda = (\gamma, d)$,
and consider $\lambda$ as
the bifurcation parameter.
The bifurcation is from the trivial solution,
$(u,v) = (0,0)$.

\subsection{Abstract Evolution Equation\label{ss-abstract}}
The system of Eqs.~(\ref{eq-uv})
defines an abstract evolution equation
for a vector-valued function
$w: [0, \infty) \to H = (L^2(\Omega))^2$,
\begin{equation}
  \frac{dw}{dt}
  =
  L_\lambda w + G_\lambda (w) , \, t > 0 ; \quad
  w(t)
  =
  \left( \begin{array}{c} u(\cdot\,, t) \\ v(\cdot\,, t) \end{array} \right) .
\Label{eq-w}
\end{equation}
Here, $L_\lambda$ a linear operator in $H$
of the form
\begin{equation}
  L_\lambda = -AD + \gamma B ,
\Label{def-L}
\end{equation}
where
$A: \dom(A) \to H$ is given by the expression
\begin{equation}
  A
  =
  - \Delta I
  =
  \left( \begin{array}{cc} -\Delta & 0 \\ 0 & -\Delta \end{array} \right) ,
\Label{def-A}
\end{equation}
on $\dom(A) = H_1 = \{ w \in (H^2(\Omega))^2 :
n\cdot \nabla w = 0 \mathrm{~on~} \partial\Omega \}$,
and
$B: H \to H$ and $D: H \to H$ are
represented by the constant matrices
\begin{equation}
  B
  =
  \left( \begin{array}{cc} f_u & f_v \\ g_u & g_v \end{array} \right) , \quad
  D
  =
  \left( \begin{array}{cc} 1 & 0 \\ 0 & d \end{array} \right) .
\Label{def-BD}
\end{equation}
In Eq.~(\ref{def-A}),
$\Delta$ denotes the Laplacian,
$H^2 (\Omega)$ is the usual Sobolev space,
and the gradient on the boundary $\partial\Omega$
of $\Omega$ is taken component-wise.

The nonlinear operator
$G_\lambda: H \to H$
is given by
\begin{equation}
  G_\lambda :
  w
  = \left( \begin{array}{c} u \\ v \end{array} \right)
  \mapsto
  \gamma
  \left( \begin{array}{c} f_1 (u,v) \\ g_1 (u,v) \end{array} \right) ,
\Label{def-G}
\end{equation}
Without loss of generality,
we assume that $G_\lambda$
can be written as the sum
of symmetric multilinear forms,
\be
  G_{\lambda}(w) = \sum _{k=2}^\infty G_{\lambda,k} (w, \ldots\,,w) ,
\Label{def-G2G3}
\ee
where $G_k$ is a symmetric $k$-linear form ($k=2,3,\ldots$).
When the $k$ arguments of $G_k$ coincide,
we write $G_k$ with a single argument,
$G_k (w) = G_k (w, \ldots\,, w)$.

The abstract evolution equation~(\ref{eq-w})
belongs to a class of equations
analyzed in detail in Ref.~\cite{mw};
the relevant results are summarized
for reference purposes in the Appendix,
Section~\ref{ss-abt}.

\section{Exchange of Stability\label{s-exchange}}
The inequalities~(\ref{ineq2-fg}) imply that
\be
  \det (B) > 0 , \quad \tr (B) < 0 .
\ee
Under these conditions,
diffusion has a destabilizing effect:
At some critical value $\lambda_0$ of $\lambda$,
an exchange of stability occurs and
the solution of Eq.~(\ref{eq-w})
bifurcates from the trivial solution.

\subsection{Eigenvalues and Eigenvectors of
$L_{\lambda}$ and $L^*_{\lambda}$\label{ss-eigen}}
The negative Laplacian $-\Delta$
on a bounded domain $\Omega \in \R^n$
with Neumann boundary conditions
is selfadjoint and positive in $L^2(\Omega)$.
Its spectrum is discrete,
consisting of eigenvalues $\rho_k$
with corresponding eigenvectors~$\varphi_k$,
\be
  - \Delta \varphi_k = \rho_k \varphi_k , \quad k = 1,2,\ldots \,,
\Label{ev-delta}
\ee
We assume that the eigenvalues are ordered,
$0 < \rho_1 \le \rho_2 \le \cdots\,$,
and that the eigenvectors~$\{\varphi_k\}_k$
form a basis in $L^2(\Omega)$.

It follows from the definition~(\ref{def-A})
that $A$ is selfadjoint and positive in $H$;
its spectrum is also discrete, consisting of
the same eigenvalues~$\rho_k$
and the eigenvectors~$\varphi_k$ once repeated.
The operator $L_\lambda$ reduces via projection
to its components on the linear span of each
eigenvector of $A$.
Let $E_k$ be the component of $L_{\lambda}$
in the eigenspace associated with
the eigenvalue $\rho_k$,
\be
  E_k (\lambda) = - \rho_k D + \gamma B , \quad k = 1,2,\ldots\,.
\Label{def-Ek}
\ee
The determinant and trace of $E_k (\lambda)$ are
\be
\begin{split}
  \det(E_k (\lambda))
  &= \gamma^2 \det(B)
  + \gamma \rho_k |g_v|
  - \rho_k d (\gamma f_u - \rho_k) , \\
  \tr(E_k (\lambda))
  &= \gamma \, \tr(B) - \rho_k (1 + d) .
\end{split}
\ee
Note that $\tr(E_k(\lambda))$ is negative
everywhere in the first quadrant and
becomes more negative as $k$ increases.

The eigenvalues of $E_k (\lambda)$
come in pairs,
\begin{equation}
  \beta_{ki} (\lambda)
   = \half
   \left(
   \tr(E_k(\lambda))
   \pm
   \left( (\tr(E_k(\lambda) )^2
             - 4\,\det(E_k(\lambda)) \right)^{1/2}
   \right) , \quad i=1,2 .
\Label{L-eigenvalues}
\end{equation}
They are either complex conjugate,
with $\Re \beta_{k1} = \Re \beta_{k2} < 0$,
or they are both real,
with $\beta_{k1} + \beta_{k2} < 0$.
We identify $\beta_{k1}$ with the upper ($+$) sign
and $\beta_{k2}$ with the lower ($-$) sign,
so $\Re \beta_{k2} \le \Re \beta_{k1}$.

The eigenvector corresponding to
the eigenvalue $\beta_{ki}$
of $L_\lambda$ is
\begin{equation}
  w_{ki}
  = \left( \begin{array}{c}
       - \gamma f_v \varphi_k \\
       (\gamma f_u - \rho_k - \beta_{ki}) \varphi_k \end{array} \right) .
\Label{L-eigenvectors}
\end{equation}
The eigenvectors $w_{k1}$ and $w_{k2}$
are linearly independent
as long as $\beta_{k1} \ne \beta_{k2}$.
The set of eigenvectors $\{w_{ki}\}_{k,i}$
forms a basis for $H$.

Note that $B$ is not symmetric;
its adjoint $B^*$ is the transpose
$B'$ of $B$.
Hence, the adjoint of $L_{\lambda}$ is
$L^*_{\lambda} = - AD + \gamma B'$,
and the adjoint of $E_k(\lambda)$ is
$E_k^*(\lambda) = - \rho_k D + \gamma B'$.
The eigenvalues of $E_k^*(\lambda)$ are
$\bar{\beta}_{ki}$, $i=1,2$,
the complex conjugates of
the eigenvalues $\beta_{ki}$
of $E_k(\lambda)$ given in
Eq.~(\ref{L-eigenvalues}).
Since the latter are either
complex conjugate or real,
the eigenvalues of $L_\lambda$
and $L_\lambda^*$ coincide.
The eigenvector corresponding
to the eigenvalue $\bar{\beta}_{ki}$
of $L_\lambda^*(\lambda)$ is
\begin{equation}
  w^*_{ki}
  = \left(\begin{array}{c}
       -\gamma g_u \varphi_k \\
     (\gamma f_u - \rho_k - \bar{\beta}_{ki}) \varphi_k\end{array} \right) .
\Label{L*-eigenvectors}
\end{equation}

\subsection{Exchange of Stability\label{ss-exchange}}
\begin{figure}[htb]
\begin{small}
\begin{center}
\setlength{\unitlength}{1mm}
\begin{picture}(75,85)%(50,55)
 \thicklines
 \put(-10,0){\vector(0,1){75}}  %vertical axis
 \put(-15, 5){\vector(1,0){95}} %horizontal axis
 %\put(-15, 5){\vector(1,0){85}} %horizontal axis
 \put(1, 0){\line(0,1){75}} %vertical asymptote at \rho_1/f_u
 %\put(21, 0){\line(0,1){75}} %vertical asymptote at \rho_2/f_u
 \put(38,0){$\gamma_1$}
 \put(40, 5){\line(0,1){2}} %tick mark at gamma_1
 \put(78,0){$\gamma$}
 %\put(68,0){$\gamma$}
 \put(-14,68){$d$}
 \put(2,0){$\rho_1/f_u$}
 %\put(2,0){$\gamma=\rho_1/f_u$}
 %\put(22,0){$\gamma=\rho_2/f_u$}
 \qbezier(2.000,75.000)(2.000,60.000)(3.000,55.000) %Curve \Lambda_1
 \qbezier(3.000,55.000)(10.000,15.000)(20.000,20.000)
 \qbezier(20.000,20,000)(25.000,20.000)(75.000,50.000)
 \qbezier(32.000,75.000)(32.000,60.000)(33.000,55.000) %Curve \Lambda_2
 \qbezier(33.000,55.000)(40.000,15.000)(50.000,20.000)
 \qbezier(50.000,20,000)(55.000,20.000)(90.000,43.000)
 \put(60,48){$\Lambda_1$}
 \put(80,42){$\Lambda_2$}
 \put(8,13){$R_1$}
 \put(15,40){$R_2$}
\end{picture}
\end{center}
\end{small}
\caption{Positive branches of $\Lambda_1$ and $\Lambda_2$.}
\label{fig1}
\end{figure}
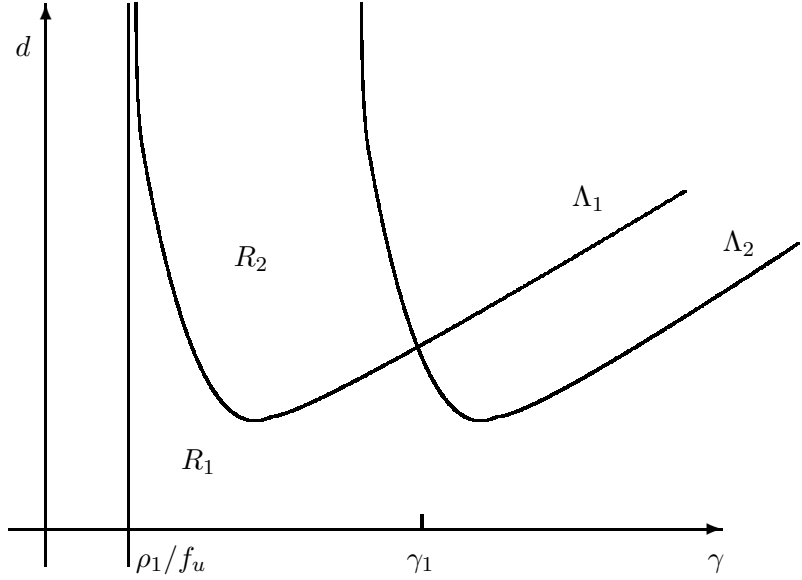

The equation $\det(E_k(\lambda)) = 0$ defines
a curve $\Lambda_k$ in the $(\gamma, d)$-plane,
\begin{equation}
  \Lambda_k
  = \{ (\gamma, d) : \det(E_k(\lambda)) = 0 \}
  = \{ (\gamma, d) : d = d_k (\gamma) \} , \quad k = 1,2, \ldots ,
\Label{Lambda-k}
\end{equation}
where
\begin{equation}
  d_k (\gamma)
  =
  \frac{\gamma^2 \det(B) + \gamma \rho_k |g_v|}
       {\rho_k (\gamma f_u - \rho_k)} .
\Label{d-k}
\end{equation}
The expression for $d_k$ can be recast in the form
\begin{equation}
  d_k (\gamma) - d^{(s)}
  =
  \frac{\rho_k |f_v g_u|}{f_u^3}
  (\gamma - \gamma_k^{(s)})^{-1}
  + \frac{\det(B)}{\rho_k f_u}
  (\gamma - \gamma_k^{(s)})^{-1} ,
\end{equation}
where
$\gamma_k^{(s)} = \rho_k/f_u$
and
$d^{(s)} = (\det(B) + |f_v g_u|)/f_u^2$.
This expression shows that
(i)~~$\Lambda_k$ is symmetric
with respect to the point $(\gamma_k^{(s)}, d^{(s)})$
(ii)~~$\Lambda_k$ has a vertical asymptote
at $\gamma = \rho_k/f_u$; and
(iii)~~$\Lambda_k$ has an oblique asymptote
with slope $\det(B)/(\rho_k f_u)$.
The symmetry point $(\gamma_k^{(s)}, d^{(s)})$
is located in the first quadrant
of the $(\gamma, d)$-plane;
$\gamma_k^{(s)}$ increases as $k$ increases,
$d^{(s)}$ is independent of $k$.
The vertical asymptote is in the right-half
of the $(\gamma, d)$-plane,
shifting to the right as $k$ increases.
The slope of the oblique asymptote
is positive, decreasing to zero as $k$ increases.
Therefore, each curve~$\Lambda_k$ has a branch
in the positive quadrant of the $(\gamma, d)$-plane.
The positive branches of the curves
$\Lambda_1$ and $\Lambda_2$
are sketched in Fig.~\ref{fig1}.

The curve $\Lambda_k$ separates the region where
$\det(E_k (\lambda)) > 0$ (below the curve)
from the region where $\det(E_k (\lambda)) < 0$
(above the curve).
We focus on the region
below the curve $\Lambda_2$,
bounded on the left by
the first vertical asymptote
at $\gamma = \rho_1/f_u$
and on the right by
$\gamma = \gamma_1$,
where the curves $\Lambda_1$
and $\Lambda_2$ cross
(see Fig.~\ref{fig1}).
The curve $\Lambda_1$ separates
this region into two subregions,
\be
\begin{split}
  R_1
  &= \{ \lambda = (\gamma, d) :
  \rho_1/f_u < \gamma < \gamma_1 , \,
  0 < d < d_1(\gamma) \} , \\
  R_2
  &= \{ \lambda = (\gamma, d) :
  \rho_1/f_u < \gamma < \gamma_1 , \,
  d_1(\gamma) < d < d_2(\gamma) \} .
\Label{def-R1R2}
\end{split}
\ee
These regions are indicated in Fig.\ref{fig1}.

\begin{lemma} \Label{l-eigenvalues}
The eigenvalues $\beta_{1i}$ ($i=1,2$)
of $L_\lambda$ satisfy the inequalities
\begin{equation}
\begin{array}{ll}
  \Re\beta_{11}(\lambda) < 0 , \, \Re\beta_{12}(\lambda) < 0 &
  \text{~if~} \lambda \in R_1 , \\
  \beta_{11}(\lambda) = 0, \, \beta_{12}(\lambda) < 0 &
  \text{~if~} \lambda \in \Lambda_1 , \\
  \beta_{11}(\lambda) > 0, \, \beta_{12}(\lambda) < 0 &
  \text{~if~} \lambda \in R_2 . \\
\end{array}
\Label{ineq-pes1}
\end{equation}
Furthermore, for $k=2,3,\ldots\,$,
\be
  \Re\beta_{k1}(\lambda) < 0, \, \Re\beta_{k2}(\lambda) < 0 \,
  \text{~if~} \lambda \in \Lambda_1  .
\Label{ineq-pes2}
\ee
\end{lemma}

\begin{proof}
In $R_1$, we have
$\tr(E_1(\lambda)) < 0$ and
$\det(E_1(\lambda) > 0$,
so $\beta_{11} (\lambda)$ and
$\beta_{12} (\lambda)$ are
either complex conjugate
with a negative real part,
or they are both real and negative.
On $\Lambda_1$,
the leading eigenvalue $\beta_{11}(\lambda)$
is zero.
Since $\tr(E_1(\lambda)) < 0$,
it must be the case that $\beta_{12} (\lambda)$
is real and negative.
In $R_2$, we have
$\tr(E_1(\lambda)) < 0$ and
$\det(E_1(\lambda) < 0$,
so $\beta_{11} (\lambda)$ and
$\beta_{12} (\lambda)$ are both real,
and they have opposite signs.

On $\Lambda_1$,
$\det (E_2(\lambda)) > 0$.
Since $\det (E_k(\lambda))$ increases with $k$,
it follows that
$\det (E_k(\lambda)) > 0$ for $k=2,3,\ldots$.
Also, $\tr (E_k(\lambda)) < 0$.
Hence, either $\beta_{k1} (\lambda)$ and
$\beta_{k2} (\lambda)$ are complex conjugate
with a negative real part,
or they are both real
and negative.
\end{proof}

The lemma implies that all eigenmodes
are stable as long as $\lambda$ is below
the curve $\Lambda_1$.
However, as soon as $\lambda$ crosses
the ``critical curve'' $\Lambda_1$,
the first unstable  eigenmode appears
and an exchange of stability occurs.

\section{Bifurcation Analysis -- One-dimensional Domain \label{s-1d}}
We first consider the bifurcation problem~(\ref{eq-w})
on a one-dimensional domain $\Omega = (0, \ell)$.
We reduce Eq.~(\ref{eq-w})
to its center-manifold representation
near a point $\lambda_0$ on
the critical curve $\Lambda_1$,
as proposed in Ref.~\cite{mw-b}
and sketched in the Appendix,
Section~\ref{ss-red}.

The eigenvalues and eigenvectors
of the negative Laplacian subject to
Neumann boundary conditions
(see Eq.~(\ref{ev-delta}))
are
\begin{equation*}
  \rho_k= k^2 (\pi /\ell)^2 , \quad
  \varphi_k (x) = \cos(k(\pi/\ell) x) , \, x \in \Omega ;
  \qquad k = 1, 2, \ldots.
\end{equation*}
The linear operator $L_\lambda$
decomposes into its components
\be
  E_k (\lambda)
  =
  - k^2 (\pi/\ell)^2 D + \gamma B , \quad k=1,2,\ldots\,,
\ee
with
\be
\begin{split}
  \det(E_k(\lambda))
  &=
  \gamma^2 \det(B)
  + \gamma |g_v| k^2 (\pi/\ell)^2
  - d k^2 (\pi/\ell)^2
  (\gamma f_u - k^2 (\pi/\ell)^2) , \\
  \tr(E_k(\lambda))
  &=
  \gamma \tr(B) - (1+d) k^2 (\pi/\ell)^2 .
\end{split}
\ee
Each $E_k$ contributes two eigenvalues,
$\beta_{k1}$ and $\beta_{k2}$,
to the spectrum of $L_\lambda$;
the expressions for $\beta_{ki}$ ($i=1,2$)
in terms of $\det(E_k(\lambda))$
and $\tr(E_k(\lambda))$
are given in Eq.~(\ref{L-eigenvalues}).
The eigenvalues of the adjoint $L_\lambda^*$
are the complex conjugates,
$\bar{\beta}_{k1}$ and $\bar{\beta}_{k2}$.
The eigenvectors of $L_\lambda$ and $L_\lambda^*$
corresponding to the eigenvalues
$\beta_{ki}$ and $\bar{\beta}_{ki}$ are
\be
  w_{ki}
  =
  \left( \begin{array}{c}
  -\gamma f_v \cos(k(\pi/\ell) x) \\
  (\gamma f_u - k^2(\pi/\ell)^2 - \beta_{ki}) \cos (k(\pi/\ell) x)
  \end{array} \right)
\Label{wki}
\ee
and
\be
  w_{ki}^*
  =
  \left( \begin{array}{c}
  -\gamma g_u \cos(k(\pi/\ell) x) \\
  (\gamma f_u - k^2(\pi/\ell)^2 - \bar{\beta}_{ki}) \cos (k(\pi/\ell) x)
  \end{array} \right) ,
\Label{wki*}
\ee
respectively.

\subsection{Center-manifold Reduction\label{ss-cmr-1d}}
We are interested in solutions of Eq.~(\ref{eq-w})
near a point $\lambda_0$ on the critical curve $\Lambda_1$.
In the region $R_1$, just below $\Lambda_1$,
both eigenvalues $\beta_{11}$ and $\beta_{12}$
are real, with $\beta_{12} < \beta_{11} < 0$.
As $\lambda$ approaches $\lambda_0$,
the leading eigenvalue $\beta_{11}$
increases and, as $\lambda$ transits
into $R_2$, $\beta_{11}$ passes
through 0 and becomes positive.
Thus, the first exchange of stability occurs.

\begin{lemma} \Label{l-reduction-1d}
Near the critical curve $\Lambda_1$,
the solution of Eq.~(\ref{eq-w})
can be expressed in the form
\be
  w = y_{11} w_{11} + z , \quad
  z = y_{12} w_{12} + \sum_{k=2}^{\infty} \sum_{i=1,2} y_{ki} w_{ki} ,
\Label{w-expand-1d}
\ee
where the coefficient $y_{11}$ of the leading term
satisfies the reduced bifurcation equation,
\begin{equation}
  \frac{dy_{11}}{dt}
  = \beta_{11} y_{11} + \alpha y_{11}^3 + o(|y_{11}|^3) .
\Label{eq4.6}
\end{equation}
The coefficient $\alpha \equiv \alpha (\lambda)$
are given explicitly in terms of the eigenfunctions
of $L_\lambda$ and $L_\lambda^*$,
\be
  \alpha (\lambda)
  =
  \alpha_2 (\lambda) + \alpha_3 (\lambda) ,
\Label{alpha-1d}
\ee
where
\begin{align*}
  \alpha_2 (\lambda)
  &=
  \frac{2}{<w_{11}, w_{11}^*>}
  \sum_{i=1,2}
  \frac{<G_2(w_{11}, w_{2i}), w_{11}^*> <G_2(w_{11}), w_{2i}^*>}
       {(2\beta_{11}-\beta_{2i}) <w_{2i},w_{2i}^*>} , \\
  \alpha_3 (\lambda)
  &=
  \frac{1} {<w_{11}, w_{11}^*>}
  <G_3(w_{11}), w_{11}^*> .
\end{align*}
Here, $< \cdot\,, \cdot >$ denotes the inner product in $H$.
(The subscript $\lambda$ on the $k$-linear forms has
been omitted.)
\end{lemma}

\begin{proof}
We look for a solution $w$ of Eq.~(\ref{eq-w})
of the form~(\ref{w-expand-1d}).
In the space spanned by the eigenvector $w_{11}$,
Eq.~(\ref{eq-w}) reduces to
\begin{equation}
\begin{split}
  <w_{11}, w_{11}^*> \frac{dy_{11}}{dt}
  &= \ <L_\lambda w, w_{11}^*> \ +\ <G_\lambda(w), w_{11}^*> \\
  &= \ \beta_{11} <w_{11}, w_{11}^*> y_{11}
  \ +\ \sum_{k=2}^\infty <G_{k} (w), w_{11}^*> .
\Label{proj.eq-1d}
\end{split}
\end{equation}
To evaluate the contributions
from the various terms in the sum,
we use the asymptotic expression for
the center-manifold function
near $\lambda_0$
given in the Appendix
(Section~\ref{ss-red}),
Theorem~\ref{th-cm-asympt},
\be
  y_{ki}
  =
  \Phi^{\lambda}_{ki} (y_{11})
  =
  \frac{<G_2(w_{11}), w_{ki}^*> y_{11}^2}
       {(2\beta_{11}-\beta_{ki}) <w_{ki}, w_{ki}^*>}
  +
  o(|y_{11}|^2) , \quad k = 2, 3, \ldots\,.
\Label{cm-1d}
\ee
The contribution from the bilinear form ($k=2$) is
\begin{align*}
  <G_2 (w), w_{11}^*>
  &=\
  <G_2 (y_{11} w_{11} + z), w_{11}^*> \\
  &=\
  <G_2 (w_{11}), w_{11}^*> y_{11}^2 \\
  &\hspace{2em}+\ 2<G_2 (w_{11}, z), w_{11}^*> y_{11} + <G_2 (z), w_{11}^*> .
\end{align*}
The first term in the right member vanishes,
because
\[
  <G_2 (w_{11}), w_{11}^*> \ = 0 .
\]
The second and third term can be evaluated
by means of the asymptotic expression~(\ref{cm-1d})
for the center manifold,
\begin{align*}
  <G_2 (w_{11}, z), w_{11}^*>
  &=
  \sum_{i=1,2} <G_2 (w_{11}, w_{2i}), w_{11}^*> y_{2i}
  +\ o(|y_{11}|^2) \\
  &=\
  \half \alpha_2 <w_{11}, w_{11}^*> y_{11}^2 + o(|y_{11}|^2) , \\
  <G_2 (z), w_{11}^*>
  &= o(|y_{11}|^3) ,
\end{align*}
where $\alpha_2$ is defined in Eq.~(\ref{alpha-1d}).
Putting it all together, we obtain the asymptotic result
\be
  <G_2(w), w_{11}^*>
  \ =\
  \alpha_2 <w_{11}, w_{11}^*> y_{11}^3 + o(|y_{11}|^3) .
\ee
The contribution from the trilinear form ($k=3$) is
\begin{equation}
\begin{split}
  <G_3 (w), w_{11}^*>
  &=\ <G_3 (w_{11}), w_{11}^*> y_{11}^3 + o(|y_{11}|^3) \\
  &=\ \alpha_3 <w_{11}, w_{11}^*> y_{11}^3 + o(|y_{11}|^3) ,
\end{split}
\end{equation}
where $\alpha_3$ is defined in Eq.~(\ref{alpha-1d}).
The higher-order forms contribute only terms of $o(|y_{11}|^3)$.
\end{proof}

\subsection{Structure of the Bifurcated Attractor\label{ss-attractor-1d}}
The results of the bifurcation analysis
for one-dimensional spatial domains
are summarized in the following theorem.

\begin{theorem}
\Label{th-1d}
$\Omega = (0,\ell)$.
\begin{itemize}
\item If $\alpha(\lambda_0) < 0$, then
the following statements are true:
\begin{enumerate}
\item
$w = 0$ is a locally asymptotically stable
equilibrium point of Eq.~(\ref{eq-w}) for
$\lambda \in R_1$ or $\lambda \in \Lambda_1$.
\item
The solution of Eq.~(\ref{eq-w})
bifurcates supercritically
from $(\lambda_0, 0)$ to
an attractor $\mA_\lambda$
as $\lambda$ crosses $\Lambda_1$
from $R_1$ into~$R_2$.
\item
There exists an open set $U_\lambda \subset H$
with $0 \in U_\lambda$ such that
the bifurcated attractor $\mA_\lambda$
attracts $U_\lambda \setminus \Gamma$ in $H$,
where $\Gamma$ is the stable manifold
of~$0$ with codimension~$1$.
\item
The attractor $\mA_\lambda$ consists
of two steady-state points,
$w_\lambda^+$ and $w_\lambda^-$,
\begin{equation}
  w_\lambda^{\pm}
  = {\pm} (\beta_{11}/|\alpha|)^{1/2} w_{11}
  + \omega_\lambda , \quad \lambda \in R_2 ,
\Label{attractor-1d}
\end{equation}
where
$\|\omega_\lambda\|_H = o (\beta_{11}^{1/2})$.
\item
There exists an $\varepsilon > 0$
and two disjoint open sets
$U_\lambda^+$ and $U_\lambda^-$ in~$H$,
with
$0 \in \partial U_\lambda^+ \cap \partial U_\lambda^-$,
such that $w_\lambda^\pm \in U_\lambda^\pm$
and
$\lim_{t \to \infty} ||w(t; w_0) - w_\lambda^\pm||_H = 0$
for any solution $w(t; w_0)$ of Eq.~(\ref{eq-w})
satisfying the initial condition
$w(0; w_0) = w_0 \in U_\lambda^\pm$
and any $\lambda$ satisfying the condition
$\mathrm{dist} (\lambda_0, \lambda) < \varepsilon$.
\end{enumerate}
\item
If $\alpha(\lambda_0) > 0$,
then the solution of Eq.~(\ref{eq-w})
bifurcates subcritically from $(\lambda_0, 0)$
to exactly two repeller points
as $\lambda$ crosses $\Lambda_1$
from $R_2$ into $R_1$.
\end{itemize}
\end{theorem}

\begin{proof}
Equation~(\ref{eq4.6}) shows that,
if $\alpha(\lambda_0) < 0$,
then $w=0$ is a locally
asymptotically stable
equilibrium point.

According to the attractor bifurcation theorem
(Section~\ref{ss-abt}, Theorem~\ref{t-abt}),
the system bifurcates at $(\lambda_0,0)$
to an attractor $\mA_{\lambda}$
as $\lambda$ transits from $R_1$ into $R_2$.

The structure of the attractor follows
from the stationary form of Eq.~(\ref{eq4.6}),
\[
  \beta_{11} y_{11} + \alpha y_{11}^3 + o(|y_{11}|^3) = 0 .
\]
The number and nature of the solutions of this equation
does not change if the terms of $o(|y_{11}|^3)$
are ignored, provided all solutions are regular
at the origin.
Thus, if $\alpha < 0$, we find
two solutions near $y=0$,
\be
  y_{11} = \pm (\beta_{11} / |\alpha|)^{1/2} + o(\beta^{1/2}) .
\ee
The last assertion of the theorem follows by time reversal.
\end{proof}

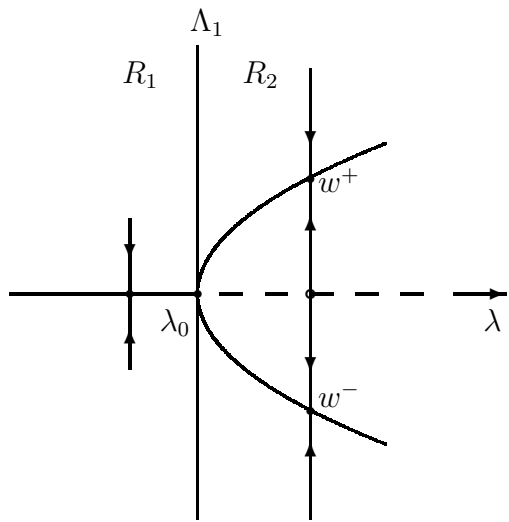
\begin{figure}[htb]
\begin{center}
 {\setlength{\unitlength}{1mm}
\begin{picture}(60,70)
  \thicklines
% XY axis
  \put(10,0){\line(0,1){63}}
  \put(-15, 30){\line(1,0){25}}
  \put(10,30){\circle*{1}}
  %\put(10, 30){\line(1,0){3}}
  \put(13, 30){\line(1,0){3}}
  \put(19, 30){\line(1,0){3}}
  \put(25, 30){\line(1,0){3}}
  \put(31, 30){\line(1,0){3}}
  \put(37, 30){\line(1,0){3}}
  \put(44,30){\vector(1,0){7}}
% Stability  Before Bifurcation
  \put(1, 20){\line(0,1){20}}
  \put(1,25){\vector(0,1){.5}}
  \put(1,35){\vector(0,-1){.5}}
  \put(1,30){\circle*{1}}
% Attractor
\qbezier(35.000,50.000)(-15.000,30.000)
         (35.000,10.000)
         \put(25, 0){\line(0,1){60}}
% Attractor after Bifurcation
    \put(25,20){\vector(0,-1){.5}}
    \put(25,40){\vector(0,1){.5}}
    \put(25,50){\vector(0,-1){.5}}
    \put(25,10){\vector(0,1){.5}}
    \put(25, 30){\circle{1}}
     \put(25, 14.45){\circle*{1}}
     \put(25, 45.3){\circle*{1}}
     \put(26, 15){$w^{-}$}
     \put(26, 43.5){$w^{+}$}
%letters
   \put(9,65){$\Lambda_1$}
   \put(0,58){$R_1$}
   \put(16,58){$R_2$}
   \put(5,25){$\lambda_0$}
   \put(48,25){$\lambda$}
\end{picture}}
\end{center}
\caption{One-dimensional domain:
Supercritical bifurcation to an attractor
$\mA = \{w^{+}, w^{-} \}$.}
\label{fig3}
\end{figure}

Theorem~\ref{th-1d} shows that,
if $\alpha(\lambda_0) < 0$,
the attractor consists of two steady-state points,
each with its own basin of attraction.
The attractor bifurcation is shown
schematically in Fig.~\ref{fig3}.
From the perspective of pattern formation,
the theorem predicts the persistence
of two types of patterns
that differ only in phase;
which of the two patterns is actually
realized depends on the initial data.

\section{Bifurcation Analysis -- Two-dimensional Domains\label{s-2d}}
Next, we consider the bifurcation problem~(\ref{eq-w})
on a two-dimensional domain
$\Omega = (0, \ell_1) \times (0, \ell_2)$.

As in the one-dimensional case,
we reduce Eq.~(\ref{eq-w})
to its center-manifold representation
near a point $\lambda_0 \in \Lambda_1$.

The eigenvalues and eigenvectors
of the negative Laplacian subject to
Neumann boundary conditions
(see Eqs.~(\ref{wki}) and (\ref{wki*}))
are
\begin{equation*}
\begin{split}
  \rho_{k_1k_2} &= k_1^2 (\pi/\ell_1)^2 + k_2^2 (\pi/\ell_2)^2 , \\
  \varphi_{k_1k_2} (x)
  &= \cos(k_1 (\pi/\ell_1) x_1) \cos(k_2 (\pi/\ell_2) x_2) , \,
  x = (x_1, x_2) \in \Omega .
\end{split}
\end{equation*}
Here, $k_1$ and $k_2$ range over all nonnegative integers
such that $|k| = k_1 + k_2 = 1, 2, \ldots$.

The eigenvalues $\beta_{k_1k_2i}$ ($i=1,2$)
and the corresponding eigenvectors of $L_\lambda$
are given in Eqs.~(\ref{L-eigenvalues})
and~(\ref{L-eigenvectors}), respectively,
where $k$ now stands for the ordered pair $(k_1,k_2)$.

The dynamics depend on the relative size of
$\ell_1$ and $\ell_2$.
On a rectangular (non-square) domain,
they are essentially the same as
on a one-dimensional domain.
For example, if $\ell_2 < \ell_1$,
then $\rho_{10} = (\pi/\ell_1)^2$
is the smallest eigenvalue of the
negative Laplacian, with corresponding
eigenvector $\varphi_{10} = \cos((\pi/\ell_1)x_1)$,
and the leading eigenvalue of $L_\lambda$
is $\beta_{101}$.
This eigenvalue is simple, and
the corresponding eigenvector is
\be
  w_{101}
  =
  \left( \begin{array}{c}
  -\gamma f_v \cos((\pi/\ell_1) x_1) \\
  (\gamma f_u - (\pi/\ell_1)^2 - \beta_{101}) \cos ((\pi/\ell_1) x_1)
  \end{array} \right) .
\ee
The center-manifold reduction leads to
a one-dimensional dynamical system
similar to Eq.~(\ref{eq4.6}).
Lemma~\ref{l-reduction-1d} and Theorem~\ref{th-1d}
apply verbatim if $\beta_{11}$ is replaced by
$\beta_{101}$ and $w_{11}$ by $w_{101}$ everywhere.

On the other hand, the dynamics become
qualitatively different if the domain
is square---that is, if
$\ell_1 = \ell_2 = \ell$
and $\Omega = (0,\ell)^2$.
The eigenvalues and eigenvectors
of $-\Delta$ on the square are
\begin{align*}
  \rho_{k_1k_2} &= (k_1^2 + k_2^2) (\pi/\ell)^2 , \\
  \varphi_{k_1k_2} (x) &= \cos(k_1(\pi/\ell)x_1) \cos(k_2(\pi/\ell)x_2) , \quad
  x = (x_1, x_2) .
\end{align*}
Note that $\rho_{k_1k_2} = \rho_{k_2k_1}$
for any pair $(k_1,k_2)$,
so the eigenvalues $\beta_{k_1k_2i}$ ($i=1,2$)
of $L_\lambda$ satisfy the same symmetry condition,
$\beta_{k_1k_2i} = \beta_{k_2k_1i}$.
To avoid notational complications,
we consider two eigenvalues, even if
they coincide because of symmetry,
as distinct and associate with each
its own eigenvector.
Thus, we associate the eigenvector
\be
  w_{k_1k_2i}
  =
  \left( \begin{array}{c}
  -\gamma f_v \varphi_{k_1k_2} \\
   (\gamma f_u - \rho_{k_1k_2} - \beta_{k_1k_2i}) \varphi_{k_1k_2}
  \end{array} \right)
\Label{wk1k2i}
\ee
with the eigenvalue $\beta_{k_1k_2i}$,
and the eigenvector
\be
  w_{k_1k_2i}^*
  =
  \left( \begin{array}{c}
  -\gamma g_u \varphi_{k_1k_2} \\
   (\gamma f_u - \rho_{k_1k_2} - \bar{\beta}_{k_1k_2i}) \varphi_{k_1k_2}
  \end{array} \right)
\Label{wk1k2i*}
\ee
with the eigenvalue $\bar{\beta}_{k_1k_2i}$,
whether $k_1$ and $k_2$ are equal or not.

\subsection{Center-manifold Reduction\label{ss-cmr-2d}}
We are again interested in values of $\lambda$
near the critical curve $\Lambda_1$,
where the first exchange of stability occurs.
The leading eigenvalues are
$\beta_{101}$ and $\beta_{011}$.
These eigenvalues coincide,
but we consider them separately,
each with its own eigenvector.
The two eigenvalues pass (together)
through 0 as $\lambda$ crosses
$\Lambda_1$ into $R_2$ from $R_1$,
at the value $\lambda = \lambda_0$.

\begin{lemma} \Label{l-reduction-2d}
Near the critical curve $\Lambda_1$,
the solution of Eq.~(\ref{eq-w})
can be expressed in the form
\be
  w = y_1 w_1 + y_2 w_2 + z , \quad
  z = \sum_{(k_1,k_2): |k|=2,3,\ldots} \sum_{i=1,2} y_{k_1k_2i} w_{k_1k_2i} ,
\Label{w-expand-2d}
\ee
where $w_1 = w_{101}$ and $w_2 = w_{011}$.
The coefficients $y_1$ and $y_2$
of the leading terms satisfy a system
of equations of the form
\begin{equation}
\begin{split}
  \frac{dy_1}{dt}
  &=
  \beta_{101} y_1 + (\alpha y_1^2 + \sigma y_2^2) y_1 + o(|y|^3) , \\
  \frac{dy_2}{dt}
  &=
  \beta_{011} y_2 + (\alpha y_2^2 + \sigma y_1^2) y_2 + o(|y|^3) ,
\Label{eq.cmr2D}
\end{split}
\end{equation}
where $\beta_{100} = \beta_{011}$.
The coefficients $\alpha \equiv \alpha(\lambda)$
and $\sigma \equiv \sigma(\lambda)$
are given explicitly in terms of the eigenfunctions
of $L_\lambda$ and $L_\lambda^*$,
\begin{equation}
  \alpha (\lambda)
  =
  \alpha_2(\lambda) + \alpha_3(\lambda) ,
\Label{alpha-2d}
\end{equation}
where
\begin{align*}
  \alpha_2 (\lambda)
  &=
  \frac{2}{<w_1, w_1^*>}
  \sum_{i=1,2}
  \frac
   {<G_2(w_1, w_{20i}), w^*_1> <G_2(w_1), w_{20i}^*>}
   {(2\beta_{101}-\beta_{20i}) <w_{20i}, w_{20i}^*>} , \\
  \alpha_3 (\lambda)
  &=
  \frac{1}{<w_1, w_1^*>}
  <G_3(w_1), w_1^*> ,
\end{align*}
and
\be
  \sigma(\lambda)
  =
  \sigma_2(\lambda) + \sigma_3(\lambda) ,
\Label{sigma-2d}
\ee
where
\begin{align*}
  \sigma_2 (\lambda)
  &=
  \frac{4}{<w_1, w_1^*>}
  \sum_{i=1,2}
  \frac{<G_2(w_2, w_{11i}), w^*_1> <G_2(w_1, w_2), w^*_{11}>}
       {(2\beta_{101}-\beta_{11i}) <w_{11i}, w_{11i}^*>} , \\
  \sigma_3 (\lambda)
  &=
  \frac {3}{<w_1, w_1^*>}
  <G_3(w_1, w_2, w_2), w_1^*> .
\end{align*}
The asymptotic estimates in Eq.~(\ref{eq.cmr2D})
are valid as $|y| = \|y_1\|+\|y_2\| \to 0$.
\end{lemma}

\begin{proof}
We look for a solution $w$ of Eq.~(\ref{eq-w})
of the form (\ref{w-expand-2d}).
In the space spanned by the eigenvectors
$w_1 = w_{101}$ and $w_2 = w_{011}$,
Eq.~(\ref{eq-w}) reduces to
\be
\begin{split}
  <w_1, w_1^*> \frac{dy_1}{dt}
  &=
  \beta_{101} <w_1, w_1^*> y_1 + \sum_{k=2}^\infty <G_k(w), w_1^*> , \\
  <w_2, w_2^*> \frac{dy_2}{dt}
  &=
  \beta_{011} <w_2, w_2^*> y_2 + \sum_{k=2}^\infty <G_k(w), w_2^*> .
\Label{proj.eq-2d}
\end{split}
\end{equation}
To evaluate the contributions from the various terms in the sums,
we again use the asymptotic expression for the center-manifold
function near $\lambda_0$ given in the Appendix
(Section~\ref{ss-red}), Theorem~\ref{th-cm-asympt},
\begin{equation}
\begin{split}
  y_{k_1k_2i}
  &= \Phi^{\lambda}_{k_1k_2i}(y_1, y_2) \\
  &=
  \frac{\sum_{j=1,2} < G_2 (w_j), w_{k_1k_2i}^*> y_j^2}
       {(2\beta_{101}-\beta_{k_1k_2i}) <w_{k_1k_2i}, w_{k_1k_2i}^*>}
    + o(|y|^2) , \quad k_1 \not= k_2 , \\
  y_{kki}
  &= \Phi^{\lambda}_{kki}(y_1, y_2)
  =
  \frac{2 <G_2 (w_1, w_2), w_{kki}^*> y_1 y_2}
       {(2\beta_{101}-\beta_{kki}) <w_{kki}, w_{kki}^*>}
    + o(|y|^2) ,
\Label{cm-2d}
\end{split}
\end{equation}
where $|y|^2 = \|y_1\|^2 + \|y_2\|^2$.

Consider the first of Eqs.~(\ref{proj.eq-2d}).
The contribution from the bilinear form is
\begin{align*}
  <G_2 (w), w_1^*>
  &=\
  \sum_{i=1,2} <G_2 (w_i), w_1^*> y_i^2 \\
  &\hspace{2em}+ 2 \sum_{i=1,2} <G_2( w_i, z), w_1^*> y_i
    + <G_2 (z), w_1^*> .
\end{align*}
The first term in the right member vanishes,
because
\[
  <G_2 (w_i), w_1^*> \ = 0 , \quad i=1,2 .
\]
The last term is asymptotically small,
\[
  <G_2 (z), w_1^*> \ = o(|y|^3) .
\]
The second term involves an infinte sum
over $(k_1,k_2)$ with $|k|=2,3,\ldots$.
Many of the coefficients are zero,
because of the specific form of
$w_1$, $w_2$, and $w_{k_1k_2j}$.
The non-zero terms can be evaluated
asymptotically by means of
the expression~(\ref{cm-2d}).
In fact, the only terms that are
non-zero and contribute to the leading-order
(cubic) terms in $y$ are those
with $i=1$ and either $(k_1, k_2) = (2, 0)$
or $(k_1, k_2) = (1, 1)$.
Asymptotic expressions for
$y_{20i}$ and $y_{11i}$ ($i=1,2$)
are given in Eq.~(\ref{cm-2d}),
where we note that only the term
with $j=1$ contributes to $y_{20i}$.

Taken together, these observations show
that the contribution from the bilinear form
is
\begin{equation}
  <G_2 (w), w_1^*>
  \ =\
  \half <w_1, w_1^*> (\alpha_2 y_1^2 + \sigma_2 y_1 y_2) y_1 + o(|y|^3) ,
\Label{proj.g2-2d}
\end{equation}
where $\alpha_2$ and $\sigma_2$ are defined in
Eqs.~(\ref{alpha-2d}) and~(\ref{sigma-2d}), respectively.

The contribution from the trilinear form is
\begin{equation}
\begin{split}
  <G_3 (w), w_1^*>
  &=
  \sum_{i=1,2} <G_3 (w_i), w_1^*> y_i^3 + o(|y|^3) \\
  &=\
  <w_1, w_1^*> (\alpha_3 y_1^2 + \sigma_3 y_2^2) y_1 + o(|y|^3) ,
\end{split}
\Label{proj.g3-2d}
\end{equation}
where $\alpha_3$ and $\sigma_3$ are defined in
Eq.~(\ref{alpha-2d}) and~(\ref{sigma-2d}), respectively.

The computations for the second of Eqs.~(\ref{proj.eq-2d})
are entirely similar.
One finds the differential equation for $y_2$
given in the statement of the lemma
with the same expressions for the coefficients
$\alpha$ and $\sigma$.
We omit the details.
\end{proof}

\subsection{Structure of the Bifurcated Attractor\label{ss-attractor-2d}}
Before proceeding to the analysis of
the structure of the bifurcated attractor,
we recall the following result,
the proof of which can be found in Ref.~\cite{mw-b}.

\begin{lemma} \Label{l-structure}
Let $y_\lambda \in \R^2$ be a solution of the evolution equation
\[
  \frac{dy}{dt} = \lambda y - G_{\lambda, k} (y) + o(|y|^k) ,
\]
where $G_{\lambda,k}$ is a symmetric $k$-linear field,
$k$ odd and $k \ge 3$, satisfying the inequalities
\[
 C_1 |y|^{k+1} \le \ < G_{\lambda,k} (y), y> \ \le C_2 |y|^{k+1}
\]
for some constants $C_2>C_1>0$, uniformly in $\lambda$.
Then $y_\lambda$ bifurcates from $(y,\lambda)=(0,0)$
to an attractor $\mA_\lambda$ which is homeomorphic to $S^1$.
Morover, one and only one of the following statements is true:
\begin{enumerate}
\item $\mA_\lambda$ is a periodic orbit;
\item $\mA_\lambda$ consists of an infinite number of singular points;
\item $\mA_\lambda$ contains at most $2(k+1)$ singular points,
which are either saddle points
or (possibly degenerate) stable nodes
or singular points with index zero.
The number of saddle points is
equal to the number of stable nodes,
and both are even ($2N$, say).
If the number of singular points is
more than $4N$
($4N+n$ say, where $4N+n \le 2(k+1)$),
then the number of singular points
with index zero is $n$ and $N+n \ge 1$.
\end{enumerate}
\end{lemma}

The results of the bifurcation analysis
for two-dimensional spatial domains
are summarized in the following theorem.

\begin{theorem}
\Label{th-2d}
$\Omega = (0,\ell)^2$.
\begin{itemize}
\item
If $\alpha(\lambda_0) < 0$ and
$\alpha(\lambda_0) + \sigma(\lambda_0) < 0$,
the following statements are true:

\begin{enumerate}
\item $w = 0$ is a locally asymptotically stable
equilibrium point of Eq.~(\ref{eq-w})
for $\lambda \in R_1$ or $\lambda \in \Lambda_1$.

\item The solution of Eq.~(\ref{eq-w})
bifurcates from $(\lambda_0, 0)$
to an attractor $\mA(\lambda)$
as $\lambda$ crosses $\Lambda_1$
from $R_1$ into $R_2$.

\item The attractor $\mA(\lambda)$ is
homeomorphic to $S^1$.
\end{enumerate}

\item If $\alpha(\lambda_0) < 0$ and
$\sigma(\lambda_0) < 0$,
the attractor $\mA(\lambda)$
consists of an infinite number
of steady-state points.

\item If $\alpha(\lambda_0) < 0$ and
$\sigma(\lambda_0) \ge 0$,
the attractor $\mA(\lambda)$
consists of exactly eight steady-state points,
which can be expressed as
\be
 w_\lambda = W_\lambda + \omega_\lambda , \quad \lambda \in R_2 ,
\Label{attractor-2d}
\ee
where $W_\lambda$ belongs to the eigenspace
corresponding to $\beta_{101}$ and
$\|\omega_\lambda\|_H =  o(\|W_\lambda\|_H)$.

\item If $\alpha(\lambda_0) > 0$ and
$\alpha(\lambda_0) + \sigma(\lambda_0) > 0$,
the solutions of Eq.~(\ref{eq-w})
bifurcate from $(\lambda_0,0)$
to a repeller $\mA(\lambda)$
as $\lambda$ transits into $R_2$.
Also, $\mA(\lambda)$ is homeomorphic to $S^1$.
\end{itemize}
\end{theorem}

\begin{proof}
Equation~(\ref{proj.eq-2d}) shows that,
if $\alpha(\lambda_0) < 0$ and
$\alpha (\lambda_0) + \sigma(\lambda_0) < 0$,
then $w=0$ is a locally asymptotically stable equilibrium point.

It follows from
Lemma~\ref{l-structure} and
the attractor bifurcation theorem~\ref{t-abt},
that the system bifurcates at $(\lambda_0,0)$
to an attractor~$\mA_\lambda$
as $\lambda$ transits from $R_1$ into $R_2$,
and that $\mA_\lambda$ is homeomorphic to $S^1$.

The structure of the bifurcated attractor is found
from the stationary form of Eq.~(\ref{eq.cmr2D}).
Ignoring the terms of $o(|y|^3)$,
we have the system of equations
\be
\begin{split}
  (\beta_{101} + \alpha y_1^2 + \sigma y_2^2) y_1 = 0 , \\
  (\beta_{101} + \alpha y_2^2 + \sigma y_1^2) y_2  = 0 .
\Label{eq4.13}
\end{split}
\ee
If $\alpha <0$,
$\alpha + \sigma <0$, and
$\sigma \ge 0$,
the system~(\ref{eq4.13}) admits
eight nonzero solutions near $y=0$,
\be
\begin{split}
  y_1 &= 0 , \quad y_2^2 = \beta_{101} / |\alpha| ; \\
  y_2 &= 0 , \quad y_1^2 = \beta_{101} / |\alpha| ; \\
  y_1^2&= y_2^2 = \beta_{101} / |\alpha + \sigma| .
\end{split}
\ee
These solutions are regular, so Eq.~(\ref{eq.cmr2D})
also has eight steady-state solutions;
they differ from the solutions of Eq.~(\ref{eq4.13})
by terms that are $o(|y|)$.

The last part of the theorem follows
by reversing time.
\end{proof}

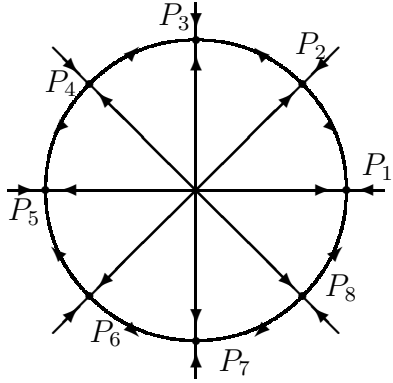
\begin{figure}[htb]
\begin{center}
\setlength{\unitlength}{1mm}
\begin{picture}(58,65)
  % Kreis M=(25,35) r=20
  \thicklines
 %Circle
  \qbezier(25.000,10.000)(33.284,10.000)
          (39.142,15.858)
  \qbezier(39.142,15.858)(45.000,21.716)
          (45.000,30.000)
  \qbezier(45.000,30.000)(45.000,38.284)
          (39.142,44.142)
  \qbezier(39.142,44.142)(33.284,50.000)
          (25.000,50.000)
  \qbezier(25.000,50.000)(16.716,50.000)
          (10.858,44.142)
  \qbezier(10.858,44.142)( 5.000,38.284)
          ( 5.000,30.000)
  \qbezier( 5.000,30.000)( 5.000,21.716)
          (10.858,15.858)
  \qbezier(10.858,15.858)(16.716,10.000)
          (25.000,10.000)
 %pointers on the circle
          \put(44,22){\vector(1,1){0.5}}
          \put(43.50,38){\vector(1,-1){0.5}}
          \put(33.284,11){\vector(-1,-1){0.5}}
          \put(33.284,48.5){\vector(-1,1){0.5}}
          \put(17,11){\vector(1,-1){0.5}}
          \put(17,48.65){\vector(1,1){0.5}}
          \put(6.5,38){\vector(-1,-1){0.5}}
          \put(6.25,22){\vector(-1,1){0.5}}
 %Inside Vectors
          \put(25, 30){\vector(1,1){13}}
          \put(25, 30){\vector(1,-1){13}}
          \put(25, 30){\vector(-1,1){13}}
          \put(25, 30){\vector(-1,-1){13}}
          \put(25, 30){\vector(0,1){18}}
          \put(25, 30){\vector(0,-1){18}}
          \put(25, 30){\vector(1,0){18}}
          \put(25, 30){\vector(-1,0){18}}
%Lines
          \put(25, 30){\line(1, 1){19}}
          \put(25, 30){\line(-1, 1){19}}
          \put(25, 30){\line(1, -1){19}}
          \put(25, 30){\line(-1, -1){19}}
          \put(25, 30){\line(0, 1){25}}
          \put(25, 30){\line(1, 0){25}}
          \put(25, 30){\line(0, -1){25}}
          \put(25, 30){\line(-1, 0){25}}
%Outside Vectors
          \put(25.000,8.000){\vector(0,1){0.5}}
          \put(41.142,13.858){\vector(-1,1){0.5}}
          \put(47.000,30.000){\vector(-1,0){0.5}}
          \put(41.142,46.142){\vector(-1,-1){0.5}}
          \put(25.000,52.000){\vector(0,-1){0.5}}
          \put(8.858,46.142){\vector(1,-1){0.5}}
          \put( 3.000,30.000){\vector(1,0){0.5}}
          \put(8.858,13.858){\vector(1,1){0.5}}
%Points & Letters
          \put(45.000,30.000){\circle*{1}}
          \put(47.000,32.000){$P_1$}
          \put(25.000,50.000){\circle*{1}}
          \put(20.000,52.000){$P_3$}
          \put( 5.000,30.000){\circle*{1}}
          \put( 0.000,26.000){$P_5$}
          \put(25.000,10.000){\circle*{1}}
          \put(28.000,6.000){$P_7$}
          \put(39.142,44.142){\circle*{1}}
          \put (38.142,48.142){$P_2$}
          \put(10.858,44.142){\circle*{1}}
          \put (4.858,43.142){$P_4$}
          \put(10.858,15.858){\circle*{1}}
          \put (10.858,9.858){$P_6$}
          \put(39.142,15.858){\circle*{1}}
          \put (42.142,15.858){$P_8$}
          %\put (20,-2){Figure3}
\end{picture}
\end{center}
\caption{Two-dimensional domain:
$S^1$-bifurcation with eight regular steady states.}
\label{fig.hetorbit}
\end{figure}

Theorem~\ref{th-2d} shows that,
if $\alpha(\lambda_0) < 0$ and
$\alpha(\lambda_0) + \sigma(\lambda_0) < 0$,
the bifurcation is an $S^1$-attractor bifurcation.
If both $\alpha(\lambda_0)$ and $\sigma(\lambda_0)$
are negative, the attractor consists of
an infinite number of steady-state points;
on the other hand, if $\alpha(\lambda_0) < 0$
and $\sigma(\lambda_0) \ge 0$,
the attractor consists of precisely eight
steady-state points.
Figure~\ref{fig.hetorbit} shows the
phase diagram on the center manifold
after bifurcation, when $\lambda$
has crossed the critical curve~$\Lambda_1$
into the region $R_2$.
The phase diagram consists of eight
steady-state points and the
heteroclinic orbits connecting them.
The odd-indexed points ($P_1$, $P_3$, $P_5$, and $P_7$)
are minimal attractors;
they correspond to striped patterns.
The even-indexed points ($P_2$, $P_4$, $P_6$ and $P_8$)
are saddle points.

\section{Examples\label{s-examples}}
We illustrate the preceding results
with two examples from the theory
of pattern formation in complex
biological structures,
namely
the Schnakenberg equation~\cite{sch}
and
the Gierer--Meinhardt equation~\cite{g-m};
see also Ref.~\cite{h-o, k-m, mpf, ni-1, ni-2, pmm, t}.

\subsection{Schnakenberg Equation}
A classic model in biological pattern formation is due to
Schnakenberg~\cite{sch},
\be
\begin{split}
  U_t &= \gamma (a - U + U^2V) + \Delta U , \\
  V_t &= \gamma (b - U^2 V) + d \Delta V ,
\Label{eq-UV-Schnak}
\end{split}
\ee
on an open bounded set $\Omega \subset \R^n$
($n=1,2)$,
with Neumann boundary conditions
and given initial conditions.
The constants $a$ and $b$ are positive;
$\gamma$ and $d$ are positive parameters.
The system admits a uniform steady state,
\be
  \left( \begin{array}{c}
  \bar{u} \\ \bar{v}
  \end{array} \right)
  =
  \left( \begin{array}{c}
  a+b \\ \frac{b}{(a+b)^2}
  \end{array} \right)
\Label{equil-Schnak}
\ee
The Schnakenberg equation is
of the type~(\ref{eq-w}),
with
\[
  B
  =
  \left( \begin{array}{cc}
    \frac{b-a}{a+b} & (a+b)^2 \\
    - \frac{2b}{a+b} & -(a+b)^2
  \end{array} \right) ,
\]
and a nonlinear term $G_\lambda$
of the form~(\ref{def-G}), with
\begin{equation*}
\begin{split}
  f_1 (u,v)
  &=
  \frac{b}{(a+b)^2} u^2 + 2(a+b)uv + u^2v , \\
  g_1 (u,v)
  &=
  - f_1 (u,v) .
\end{split}
\end{equation*}
The conditions~(\ref{ineq1-fg}) and~(\ref{ineq2-fg})
are satisfied if
\[
  a<b, \quad b-a<(a+b)^3 .
\]

\subsubsection{One-dimensional Domain}
$\Omega = (0,1)$.

An evaluation of the inner products
in Eq.~(\ref{alpha-1d})
with the \texttt{MAPLE}
software package yields
the expression
\[
  \alpha(\lambda)
  =
  \frac{s_1 + s_2 + s_3}
       {-2\gamma^2 b(a+b) + \left( \gamma \frac{b-a}{a+b} - \rho_1 \right)^2} , \quad
  \lambda \in \Lambda_1 ,
\]
where
\begin{equation*}
\begin{split}
  s_i
  =\,
  & \half \gamma^4 (\gamma + \rho_1) (a+b)^4
  ((5\rho_1 + \beta_{2i})(a+b) + \gamma (2a-b)) \\
  & \times
  \frac {(\gamma (2a-b) + 2\rho_1 (a+b))(4\rho_1 + \gamma + \beta_{2i})}
  {\beta_{2i} (2\gamma^2 b(a+b) - (\gamma \frac{b-a}{a+b} - \rho_2 - \beta_{2i})^2)} ,
  \quad i=1,2 , \\
  s_3
  =\,
  & \textstyle{\frac{3}{4}}
  \gamma^3 (\gamma + \rho_1) (a+b)^3
  \left(\gamma (b-a) - \rho_1 (a+b) \right) .
\end{split}
\end{equation*}
Note that
$\alpha(\lambda)$ depends only on $\gamma$
if $\lambda \in \Lambda_1$;
we use the short-hand notation
$\alpha(\gamma) \equiv \alpha(\gamma, d_1(\gamma))$.

\subsubsection{Two-dimensional Domain}
$\Omega = (0,1)^2$.

An evaluation of the inner products
in Eqs.~(\ref{alpha-2d}) and~(\ref{sigma-2d})
with the \texttt{MAPLE} software package
yields the expressions
\[
  \alpha(\lambda)
  =
  \frac{s_1 + s_2 + s_3}
       {-2\gamma^2 b(a+b) + \left( \gamma \frac{b-a}{a+b} - \rho_{01} \right)^2} , \quad
  \lambda \in \Lambda_1 ,
\]
and
\[
  \sigma(\lambda)
  =
  \frac{s^1 + s^2 + s^3}
       {-2\gamma^2 b(a+b) + (\gamma \frac{b-a}{a+b} - \rho_{01})^2} , \quad
  \lambda \in \Lambda_1 ,
\]
where
\begin{equation*}
\begin{split}
  s_i
  =\,
  & \half \gamma^4 (\gamma + \rho_1) (a+b)^4
  ((5 \rho_1 + \beta_{20i}) (a+b) + \gamma (2a-b)) \\
  & \times
  \frac{(\gamma (2a-b) + 2\rho_1(a+b)) (4\rho_1 + \gamma +\beta_{20i})}
  {\beta_{20i} (2 \gamma^2 b (a+b) - (\gamma \frac{b-a}{b+a} - 4 \rho_{10}
   - \beta_{20i})^2)} , \quad i=1,2 , \\
  s_3
  =\,
  & \textstyle\frac34 \gamma^3 (\gamma+\rho_1) (a+b)^3
  (\gamma (b-a) - \rho_1 (a+b) ) ; \\
  s^i
  =\,
  & \half \gamma^4 (a+b)^4 (\gamma + \rho_1)
  ((2a-b) \gamma + (\beta_{11i}+3\rho_1) (a+b)) \\
  & \times
  \frac{((2a-b) \gamma + 2(a+b) \rho_1) (\gamma +2\rho_1+\beta_{11i})}
       {\beta_{02i}(2 \gamma^2 b (a+b)
  - (\gamma \frac{b-a}{a+b} - 2 \rho_{01} - \beta_{11i})^2)} , \quad i=1,2 , \\
  s^3
  =\,
  & \textstyle\frac32 \gamma^3 (\gamma + \rho_{01}) (a+b)^3
  (\gamma (b-a) - \rho_{01} (a+b)) .
\end{split}
\end{equation*}

\subsubsection{Numerical Results}
\begin{center}
\begin{figure}[htb]
\includegraphics[height=2.5in]{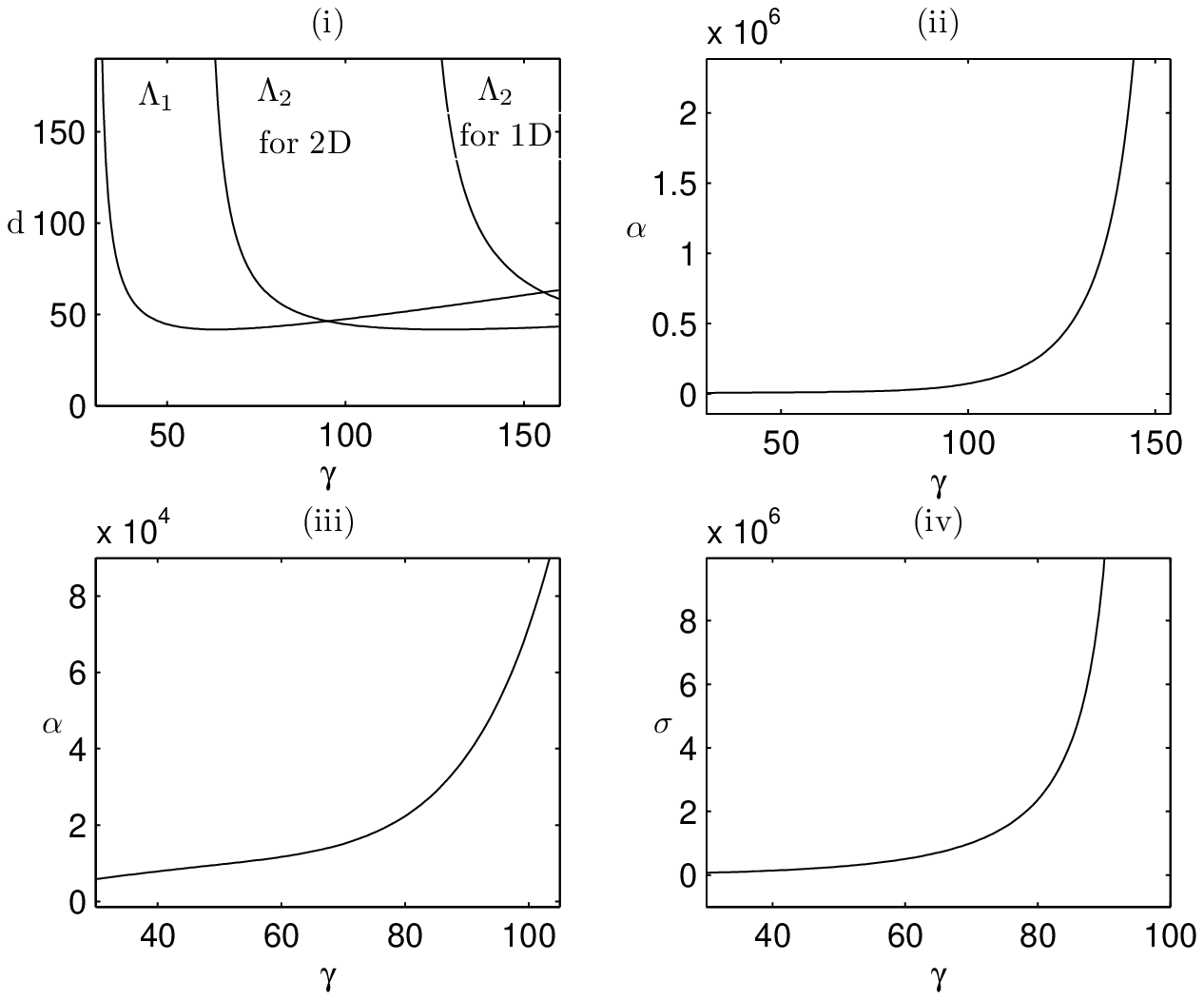}
\caption{Schnakenberg equation with
$a=\onethird$, $b =\twothirds$.
(i)~Positive branches of $\Lambda_1$
and $\Lambda_2$ in one and two dimensions;
(ii)~Graph of $\alpha$ in the one-dimensional case;
(iii)~Graph of $\alpha$ in the two-dimensional case;
(iv)~Graph of $\sigma$ in the two-dimensional case.
\label{fig4}}
\end{figure}
\end{center}

\begin{center}
\begin{figure}[htb]
\includegraphics[height=2.5in]{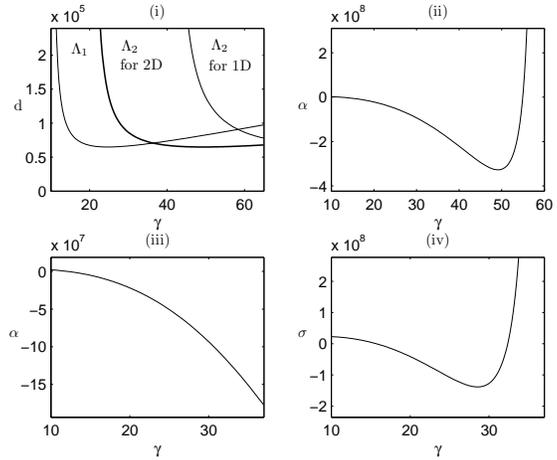}
\caption{Schnakenberg equation with
$a=2$, $b =100$.
(i)~Positive branches of $\Lambda_1$
and $\Lambda_2$ in one and two dimensions;
(ii)~Graph of $\alpha$ in the one-dimensional case;
(iii)~Graph of $\alpha$ in the two-dimensional case;
(iv)~Graph of $\sigma$ in the two-dimensional case.
\label{fig5}}
\end{figure}
\end{center}

Numerical results are given for
$a=\onethird$, $b =\twothirds$
in Fig.~\ref{fig4},
and for
$a=2$ and $b=100$
in Fig.~\ref{fig5}.
In the former case, there is no bifurcation;
in the latter, there is a pitchfork bifurcation
at $\lambda_0=(\gamma_0,d(\gamma_0))$.

\subsection{Gierer--Meinhardt Equation}
Another model for the formation of Turing patterns
was proposed by Gierer and Meinhardt~\cite{g-m},
\be
\begin{split}
  U_t &= \gamma (a - bU + U^2/V) + \Delta U , \\
  V_t &= \gamma (U^2 - V) + d \Delta V,
\Label{eq-UV-GM}
\end{split}
\ee
on an open bounded set $\Omega \subset \R^n$ ($n=1,2$),
with Neumann boundary conditions
and given initial conditions.
The constants $a$ and $b$ are positive.
The system admits a steady state,
\be
  \left( \begin{array}{c}
  \bar{u} \\ \bar{v}
  \end{array} \right)
  =
  \left( \begin{array}{c}
  \frac{a+1}{b} \\ \frac{(a+1)^2}{b^2}
  \end{array} \right)
\Label{equil-GM}
\ee
The Gierer--Meinhardt system is an equation
of the type~(\ref{eq-w}) with
\[
  B
  =
  \left( \begin{array}{cc}
  \frac{(1-a)b}{1+a} & - \frac{b^2}{(1+a)^2} \\
  \frac{2(1+a)}{b} & -1
  \end{array} \right) .
\]
The nonlinear term $G_\lambda$ is obtained
by expanding around the steady-state solution,
\begin{equation*}
\begin{split}
  f_1 (u,v)
  &=
  \frac{b^2}{(1+a)^2}
  \left(
  u^2 - \frac{2b}{1+a} uv \right. \\
  &\left.\hspace{2em}+ \frac{b^2}{(1+a)^2}
  \left(
  v^2 - u^2v + \frac{2b}{1+a} uv^2 + \cdots
  \right)
  \right) , \\
  g_1 (u,v)
  &= u^2 .
\end{split}
\end{equation*}
The conditions~(\ref{ineq1-fg}) and~(\ref{ineq2-fg})
are satisfied if
\[
  a<1, \quad b<\frac{1+a}{1-a} .
\]

\subsubsection{One-dimensional Domain}
$\Omega=(0,1)$.

An evaluation of the inner products
in Eq.~(\ref{alpha-1d})
with the \texttt{MAPLE}
software package yields
the expression
\[
  \alpha(\lambda)
  =
  \frac{s_1+s_2+s_3}{-\gamma^2 \frac{2b}{1+a}
   + \left( \gamma \frac{(1-a)b}{1+a} - \rho_1 \right)^2} ,
\]
where
\begin{equation*}
\begin{split}
  s_i
  =\,
  & - \half \gamma^4 b^6 \\
  &\times
  \frac {\gamma^2 (2a-1)b^2 + \gamma b (\rho_1 +2 a ( \beta_{2i} + 5 \rho_1))
  +2 \rho_1 (1+a) ( \beta_{2i} + 4 \rho_1)}
  {(1+a)^8} \\
  &\times
  \frac{\gamma^2 (2a-1)b^2 + \gamma b ( 4 \rho_1(1+a)+ \beta_{2i})
  +2 \rho_1^2 (1+a)}
  {\beta_{2i} \left( - \gamma^2 \frac{2b}{1+a}
   + \left( \gamma \frac{(1-a)b}{1+a} - \rho_1 - \beta_{2i} \right)^2
   \right)} , \quad i=1,2, \\
  s_3
  =\,
  & {\textstyle{\frac32}}
  \frac{\gamma^2 b^5 (\gamma (1-a)b - \rho_1(1+a)) (\gamma ab +\rho_1 (1+a))^2}
  {(1+a)^8} .
\end{split}
\end{equation*}

\subsubsection{Two-dimensional Domain}
$\Omega = (0,1)^2$.

An evaluation of the inner products
in Eqs.~(\ref{alpha-2d}) and~(\ref{sigma-2d})
with the \texttt{MAPLE} software package
yields the expressions
\[
  \alpha(\lambda)
  =
  \frac{s_1+s_2+s_3}{-\gamma^2 \frac{2b}{1+a}
   + \left( \gamma \frac{(1-a)b}{1+a} - \rho_{10} \right)^2} ,
\]
and
\[
  \sigma(\lambda)
  =
  \frac{s^1+s^2+s^3}{-\gamma^2 \frac{2b}{1+a}
   + \left( \gamma \frac{(1-a)b}{1+a} - \rho_{10} \right)^2} ,
\]
where
\begin{equation*}
\begin{split}
  s_i
  =\,
  & - \half \gamma^4 b^6 \\
  &\times
  \frac {\gamma^2 b^2
  + \gamma b (\rho_{10} + 2a (\beta_{20i}+5\rho_{10}))
  +2\rho_{10} (1+a) (\beta_{20i}+4\rho_{10})} {(1+a)^8} \\
  &\times
  \frac {\gamma^2 (2a-1) b^2
  + \gamma b (4\rho_{10} (1+a) + \beta_{20i}) +2\rho_{10}^2 (1+a)}
   {\beta_{20i}
    \left( - \gamma^2 \frac{2b}{1+a}
   + \left( \gamma \frac{(1-a)b}{1+a} - \rho_{10} - \beta_{20i} \right)^2
   \right)} , \quad i=1,2, \\
  s_3
  =\,
  & {\textstyle\frac32}
  \frac {\gamma^2 b^5 ( \gamma (1-a)b - \rho_{10} (1+a))
  (\gamma ab +\rho_{10} (1+a))^2}{ (1+a)^8} ; \\
  s^i
  =\,
  & -\half \gamma^4 b^6 \\
  &\times
  \frac{\gamma^2 (2a-1)b^2 + \gamma b (\rho_{10}+2a (\beta_{11i}+3\rho_{10}))
  + 2\rho_{10} (1+a) (\beta_{11i}+2\rho_{10})}
  {(1+a)^8} \\
  &\times
  \frac{\gamma^2 (2a-1)b^2 + \gamma  b (\beta_{11i} + 2\rho_{10} (2a+1) )
  +2\rho_{10}^2 (1+a)}
  {\beta_{11i} \left( - \gamma^2 \frac{2b}{1+a}
   + \left( \gamma \frac{(1-a)b}{1+a} - \rho_{10} - \beta_{11i} \right)^2
   \right)} , \quad i=1,2 , \\
  s^3
  =\,
  & 3 \frac {\gamma^2 b^5 (\gamma (1-a)b - \rho_{10} (1+a))
  (\gamma ab +\rho_{10} (1+a))^2}{(1+a)^8} .
\end{split}
\end{equation*}

\subsubsection{Numerical Results}
\begin{center}
\begin{figure}[htb]
\includegraphics[height=2.5in]{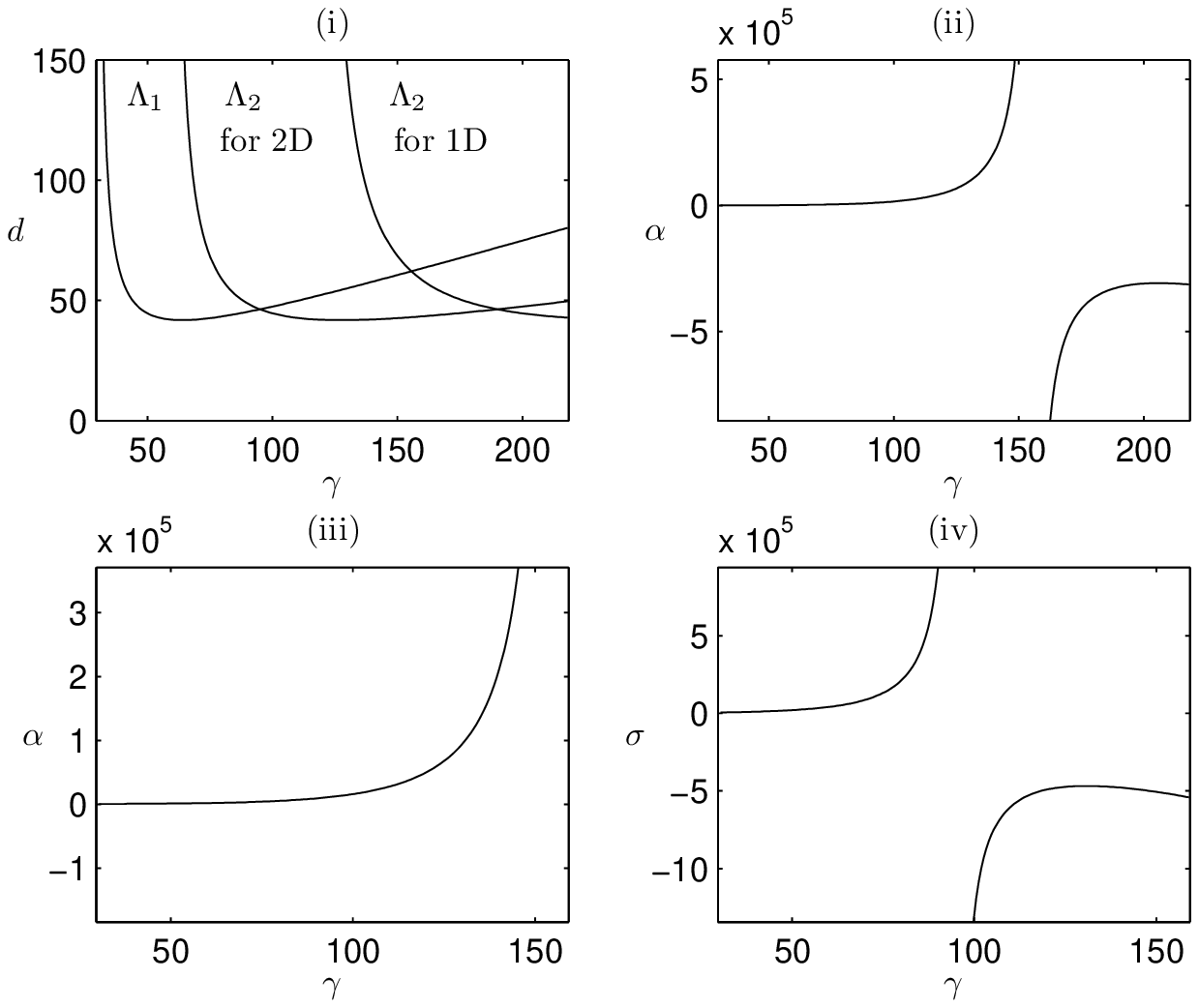}
\caption{Gierer--Meinhardt equation with $a=\frac12$, $b=1$.
(i)~Positive branches of $\Lambda_1$ and $\Lambda_2$ in one and two dimensions;
(ii)~Graph of $\alpha$ in the one-dimensional case;
(iii)~Graph of $\alpha$ in the two-dimensional case;
(iv)~Graph of $\sigma$ in the two-dimensional case.}
\label{fig6}
\end{figure}
\end{center}

\begin{center}
\begin{figure}[htb]
\includegraphics[height=2.5in]{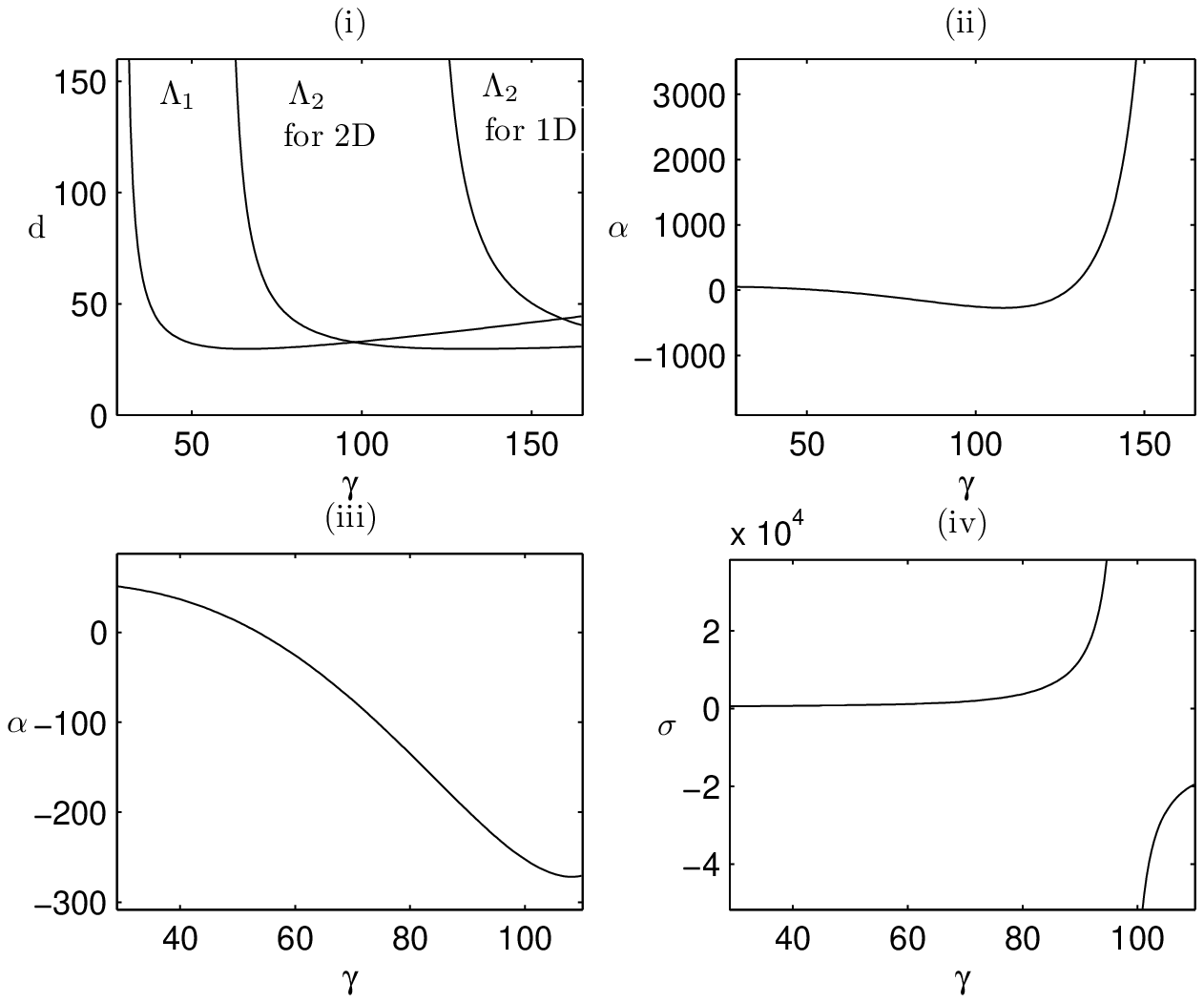}
\caption{Gierer--Meinhardt equation with $a=\onethird$, $b=\twothirds$.
(i)~Positive branches of $\Lambda_1$ and $\Lambda_2$ in one and two dimensions;
(ii)~Graph of $\alpha$ in the one-dimensional case;
(iii)~Graph of $\alpha$ in the two-dimensional case;
(iv)~Graph of $\sigma$ in the two-dimensional case.}
\label{fig7}
\end{figure}
\end{center}

Numerical results are given
for
$a=\half$, $b =1$
(Fig.~\ref{fig6})
and
$a=\onethird$ and $b=\twothirds$
(Fig.~\ref{fig7}).
In the former case, there is no bifurcation;
in the latter, there is a bifurcation
at $\lambda_0=(\gamma_0,d(\gamma_0))$.

\section{Conclusions\label{s-conclusions}}
In this paper we considered the evolution
of an activator--inhibitor system
consisting of two morphogens
on a bounded domain subject to
no-flux boundary conditions.
Assuming that the system admits
a uniform steady-state solution,
which is stable in the absence of diffusion,
we focused on solutions that bifurcate
from this uniform steady-state solution.
The bifurcation parameter $\lambda$
represented both $\gamma$,
the ratio of the characteristic times
for chemical reaction and diffusion,
and $d$, the ratio of the diffusion coefficients
of the two competing species (activator
and inhibitor).
We showed that, for such a system,
there exists a critical curve $\Lambda_1$
in parameter space such that,
as $\lambda$ crosses $\Lambda_1$,
a bifurcation occurs (Lemma~\ref{l-eigenvalues}).

While a linear analysis around the uniform
steady state suffices to obtain information
about the \emph{formation} of patterns,
a nonlinear analysis is needed to gain insight
into the long-time \emph{asymptotic behavior}
of the solutions after bifurcation.
This issue is intimately connected
with the long-term persistence of patterns.

In this paper we used the theory
of attractor bifurcation,
in combination with a
center-manifold reduction,
to analyze the long-time dynamics
of bifurcated solutions.
We considered two cases:
diffusion on a (bounded) interval
or a (non-square) rectangle,
and diffusion on a square domain.
In the former case, we showed that
a bifurcation occurs as $\lambda$
crosses a critical curve $\Lambda_1$,
and the bifurcation is
a pitchfork bifurcation.
Theorem~\ref{th-1d} gives an explicit condition
for the existence of an attractor.
The attractor consists of exactly two steady-state points,
each with its own basin of attraction.
The two steady states correspond to patterns
that differ only in phase;
which of them is eventually realized
depends on the initial conditions.
Essentially the same conclusion holds
in the case of diffusion on a rectangular
(that is, non-square) domain;
in particular, roll patterns emerge
as a result of the bifurcation.

In the case of diffusion on a square domain,
the dynamics are qualitatively different.
Theorem~\ref{th-2d} gives explicit conditions
for the existence of an $S^1$-bifurcation.
The bifurcated object consists of either
an infinite number of steady states,
or exactly eight regular steady-state points
with heteroclinic orbits connecting them.
Thus, two types of patterns may arise;
for example, in the formation of animal coat patterns,
we might expect stripe patterns or spot patterns.

Thus, in both the one- and two-dimensional case
we have given a complete characterization
of the bifurcated attractor and, therefore,
of the long-time asymptotic dynamics
of the bifurcated objects.

\appendix

\section{Attractor Bifurcation and Reduction Methods\label{s-appendix}}
In this appendix,
we summarize the attractor bifurcation theory of Ref.~\cite{mw}
and the reduction methods introduced in Ref.~\cite{mw-b}.
The functional framework is that of two Hilbert spaces,
$H$ and $H_1$, where $H_1$ is dense in $H$ and
the inclusion $H_1 \hookrightarrow H$ is compact.

\subsection{Attractor Bifurcation Theorem\label{ss-abt}}
Let $A : H_1 \to H$ be a linear homeomorphism,
and let $L_\lambda : H_1 \to H$ be
a compact perturbation of~$A$
which depends continuously on
a real parameter~$\lambda$,
\begin{equation}
  L_\lambda = - A + B_\lambda .
\end{equation}
The operator $L_\lambda$ is sectorial;
it generates an analytic semigroup
$S_\lambda (t) =
\{e^{tL_\lambda}\}_{t\geq 0}$.
Fractional powers $L_\lambda^\alpha$
are defined for all $\alpha \in [0,1]$;
the domain of $L_\lambda^\alpha$ is
$\dom (L_\lambda^\alpha) = H_\alpha$,
where $H_0 = H$ and
$H_{\alpha_1} \subset H_{\alpha_2}$
if $\alpha_2 < \alpha_1$.

Let $G_\lambda : H_\alpha \to H$
be a nonlinear $C^r$-bounded map
($r \ge 1$) for some $\alpha \in [0,1)$,
which depends continuously on $\lambda$
and satisfies the asymptotic estimate
\begin{equation}
  G_\lambda (w) = o (\|w\|_{H_\alpha}) ,
\Label{eq2.4}
\end{equation}
as $|w| \to 0$, uniformly in $\lambda$.

Let
$w_\lambda (t; w_0) \in H$
be a solution of the
initial value problem
\be
  \frac{dw}{dt}
  = L_\lambda w + G_\lambda (w) , \; t > 0 ; \quad
  w(0) = w_0 ,
\Label{eq.ev}
\ee
where $w_0 \in H$ is given.
In terms of $S_\lambda$, we have
\begin{equation}
  w_\lambda(t; w_0) = S_\lambda (t) w_0
  + \int_0^t S_\lambda (t-s) G_\lambda (w_\lambda(s; w_0)) ds , \quad t\geq 0 .
\end{equation}

\begin{definition}
\Label{df2.1}
(i)~~A set $\Sigma \subset H$ is a
\emph{(positive) invariant set} of
Eq.~(\ref{eq.ev})
if $S_\lambda (t) \Sigma = \Sigma$
for any $t\geq 0$.
(ii)~~An invariant set $\Sigma \subset H$
of Eq.~(\ref{eq.ev}) is an
\emph{attractor} if $\Sigma$ is compact
and there exists a neighborhood $U \subset H$
of $\Sigma$ such that
\[
 \lim_{t \to \infty} \mathrm{dist}_H (w_\lambda (t; w_0), \Sigma) = 0
\]
for any $w_0 \in U$.
(iii)~~The largest open set $U$ satisfying
the above condition is the
\emph{basin of attraction} of $\Sigma$.
\end{definition}

\begin{definition}
\Label{df2.2}
(i)~~A solution $(w_\lambda, \lambda)$ of Eq.~(\ref{eq.ev})
\emph{bifurcates} from $(0, \lambda_0)$
if there exists a sequence of invariant sets
$\{\Omega_n\}$ of Eq.~(\ref{eq.ev})
with $0 \notin \Omega_n$ such that
\[
 \lim_{n \to \infty} \max_{w \in \Omega_n} |w| = 0 , \quad
 \lim_{n \to \infty} \mathrm{dist}(\lambda_n ,\lambda_0)=0 .
\]
(ii)~~If the invariant sets $\Omega_n$ are attractors of
Eq.~(\ref{eq.ev}), then the bifurcation is called an
\emph{attractor bifurcation}.
(iii)~~If the invariant sets $\Omega_n$ are attractors
and are homotopy equivalent to an $m$-dimensional sphere $S^m$,
then the bifurcation is called an
\emph{$S^m$-attractor bifurcation}.
\end{definition}

The theory of Ref.~\cite{mw} is developed
under the conditions that
(i)~the spectrum of $A$ is discrete
and consists of positive eigenvalues~$\rho_k$
with corresponding eigenvectors~$\varphi_k \in H_1$,
\begin{equation}
  A \varphi_k = \rho_k \varphi_k,  \quad k=1,2,\ldots ,
\Label{eq2.5}
\end{equation}
where $0 < \rho_1 \le \rho_2 \le \cdots \to \infty$;
(ii)~the vectors $\{\varphi_k\}_k$ form an orthogonal basis of~$H$;
and
(iii)~there exists a constant $\theta \in (0,1)$
such that $B_\lambda : H_\theta \to H$
is bounded uniformly in $\lambda$.

The eigenvalues of $L_\lambda$ are
$\beta_k (\lambda)$, $k=1,2,\ldots$.
The assumption is that there is a critical
curve $\Lambda_m$ that separates the region $R_m$
where (the real parts of) the first $m$ eigenvalues
are negative from the region $R_{m+1}$
where (the real parts of) the first $m$ eigenvalues
are positive,
and an exchange of stanility occurs if
$ \lambda = \lambda_0 \in \Lambda_m$.

\begin{theorem}[Attractor Bifurcation Theorem~\cite{mw}]
\Label{t-abt}
Let $\beta_k (\lambda)$, $k=1,2,\ldots\,$,
be the eigenvalues of $L_\lambda$
(counting multiplicity).
Suppose that, at some critical value
$\lambda = \lambda_0$,
the first $m$ eigenvalues
$\beta_1(\lambda)$ through $\beta_m (\lambda)$
cross the imaginary axis into the
right half of the complex plane,
while the remaining eigenvalues,
$\beta_k(\lambda)$ for $k=m+1, m+2, \ldots$,
remain in the left half of the complex plane.
Let $E_1$ be the eigenspace
of $L_\lambda$ at $\lambda_0$,
\[
 E_1 = \bigcup_{k=1}^{m}
 \left\{ w \in H_1 : (L_{\lambda_0} - \beta_k(\lambda_0))^i w = 0 , \,
  i = 1, 2, \ldots \right\} ,
\]
and let $w=0$ be a locally asymptotically stable equilibrium point
of Eq.~(\ref{eq.ev}) at $\lambda = \lambda_0$.
Then
\begin{enumerate}
\item
Eq.~(\ref{eq.ev}) bifurcates
from $(w,\lambda) = (0,\lambda_0)$
to an attractor $\mA_\lambda$
as $\lambda$ transits from $R_1$ into $R_2$,
$\dim \mA_\lambda \in [m-1, m]$, and
$\mA_\lambda$ is connected if $m>1$;
\item
$\mA_\lambda$ is the limit of a sequence
of nested $m$-dimensional annuli $M_i$
with $M_{i+1}\subset M_i$;
in particular,
if $\mA_\lambda$ is a finite simplicial complex,
then $\mA_\lambda$ has the homotopy type of $S^{m-1}$;
\item
For any $w_\lambda \in \mA_\lambda$,
$w_\lambda$ can be expressed as
\[
 w_\lambda = W_\lambda + z_\lambda , \quad
 W_\lambda\in E_1 , \quad
 z_\lambda = o(\|W_\lambda\|_H) ;
\]
\item
There is an open set $U\subset H$ with $0\in U$
such that $\mA_\lambda$ attracts $U \setminus \Gamma$,
where $\Gamma$ is the stable manifold of $w=0$
with codimension $m$.
\end{enumerate}
\end{theorem}

\subsection{Reduction Method\label{ss-red}}
A useful tool in the study of bifurcation problems
is the reduction of the equation to its local
center manifold.
The idea is to project the equation
to a finite-dimensional space
after a change of basis;
details can be found in~\cite{mw-b}.

Let $\lambda$ be close to a critical value $\lambda_0$.
Suppose that the spaces $H$ and $H_1$ are decomposed,
\begin{equation}
      H = E_1 \oplus E_2 , \quad
      H_1 = \widetilde{E}_1 \oplus \widetilde{E}_2 ,
\Label{cm.1}
\end{equation}
where $E_1$ and $E_2$ are invariant subspaces of $L_{\lambda}$,
$E_1$ is finite dimensional,
$\widetilde{E}_1 = E_1$, and
$\widetilde{E}_2$ is the closure of $E_2$ in $H$.
The decomposition (\ref{cm.1}) reduces $L_{\lambda}$,
\be
  L_{\lambda} = \mathcal{L}_{\lambda,1} \oplus \mathcal{L}_{\lambda,2} ,
\ee
where
$\mathcal{L}_{\lambda,1} = L_{\lambda}|_{E_1} : E_1 \to \widetilde{E}_1$
and
$\mathcal{L}_{\lambda,2} = L_{\lambda}|_{E_2} : E_2 \to \widetilde{E}_2$.
If the decomposition is such that
the real parts of the eigenvalues of
$\mathcal{L}_{\lambda,1}$ are nonnegative
at $\lambda=\lambda_0$,
while those of $\mathcal{L}_{\lambda,1}$ are negative,
then the solution $w_\lambda$ of Eq.~(\ref{eq.ev})
can be written as
\be
  w = W + z , \quad W \in E_1 , \, z \in E_2 ,
\ee
where $W$ and $z$ satisfy the system of equations
\begin{equation}
\Label{cm.eq}
\begin{split}
  \frac{dW}{dt} &= \mathcal{L}_{\lambda,1} W + \mathcal{G}_{\lambda,1} (W,z) , \\
  \frac{dz}{dt} &= \mathcal{L}_{\lambda,2} z + \mathcal{G}_{\lambda,2}(W,z) ,
\end{split}
\end{equation}
with
$\mathcal{G}_{\lambda,i} = P_i G_{\lambda}$,
$P_i : H \to \widetilde{E}_i$ being the canonical projection.

By the classical center-manifold theorem
(see, for example, Refs.~\cite{h,t}),
there exist,
for all $\lambda$ sufficiently close to $\lambda_0$,
a neighborhood $U_{\lambda} \subset E_1$ of $W = 0$
and a $C^1$ center-manifold function
$\Phi^\lambda : U_{\lambda} \to E_1$,
which depends continuously on $\lambda$,
such that the dynamics of Eq.~(\ref{eq.ev})
are described completely by the dynamics of
the finite-dimensional system
\[
  \frac{dW}{dt}
  =
  \mathcal{L}_{\lambda,1} W
  + \mathcal{G}_{\lambda,1}(W,\Phi^\lambda(W)), \quad
  W \in U_{\lambda} \subset E_1 .
\]
The following theorem gives an asymptotic approximation
for $\Phi^\lambda$ as $\lambda \to \lambda_0$
or, alternatively, as $W \to 0$.
The proof of the theorem is given in Ref.~\cite{mw-b}.

\begin{theorem}
Assume that $G_\lambda$,
the nonlinear part of Eq.~(\ref{eq.ev}),
is $C^\infty$,
$G_\lambda (w) = \sum_{k=p}^{\infty} G_{\lambda,k} (w)$
for some $p\ge2$,
where $G_{\lambda,k} (w) = G_{\lambda,k} (w,\ldots,w)$,
and $G_{\lambda,k} : H_1 \times\ldots\times H_1 \to H$
is a $k$-linear map.
Then, under the conditions of Theorem~\ref{t-abt},
the center-manifold function $\Phi^\lambda$
can be expressed as
\begin{align*}
  \Phi^{\lambda}(W)
  &=
  (-\mathcal{L}_{\lambda,2})^{-1} P_2 G_{\lambda,p} (W)
  + O(|\Re\beta(\lambda)| \cdot ||W||^p) \\
  &\hspace{2em}\mbox{}+ o(||W||^p) ,
  \quad W \in E_1 , \; \lambda \to \lambda_0 ,
\end{align*}
where $\beta = (\beta_1, \ldots\,, \beta_m)$.
\Label{th-cm-asympt}
\end{theorem}

\subsection*{Acknowledgments}
The authors thank James Glazier (Indiana University), T.~J.~Kaper (Boston University)
and A.~Doelman (CWI, Amsterdam)
for helpful discussions on the subject of pattern formation
and for providing several of the references.

The work of H.G.K.~was supported in part by the
National Science Foundation under Award No.~DMS-0549430-001 and by the
U.S.~Department of Energy under Contract No.~DE-AC02-06CH11357.
The work of S.W.\ and M.Y.\ was supported in part by the
Office of Naval Research under contract N00014-05-1-0218
and by the National Science Foundation
under contract DMS-0605067.

\vfill
\hfill
\fbox{
\parbox{3in}{\footnotesize{
The submitted manuscript has been created in part by
the UChicago Argonne, LLC, Operator of Argonne National
Laboratory ("Argonne") under Contract No. DE-AC02-06CH11357
with the U.S. Department of Energy.
The U.S. Government retains for itself, and others acting
on its behalf, a paid-up, nonexclusive, irrevocable worldwide
license in said article to reproduce, prepare derivative works,
distribute copies to the public, and perform publicly and
display publicly, by or on behalf of the Government.
}}}


\begin{thebibliography}{99}

\bibitem{bms}
\textsc{Benson, D.~L., P.~K.~Maini, and J.~A.~Sherratt},
\textit{Unravelling the Turing bifurcation using
spatially varying diffusion coefficients},
J.~Math.\ Biol.~37, 381--417, 1998.

\bibitem{bvhs}
\textsc{Bollerman, P., A.~van Harten, and G.~Schneider},
\textit{On the Justification of the G-L Approximation}
in:
\textsl{Nonlinear Dynamics and Pattern Formation in the Natural Environment},
A.~Doelman and A.~van Harten (eds.), Longman, 1995, pp.~20--36.

\bibitem{cross-h}
\textsc{Cross, M.~C. and P.~C.~Hohenberg}
\textit{Pattern formation outside of equilibrium}
Rev.\ Mod.\ Phys.~65, 851--1112, 1993.

%\bibitem{e}
%{\sc G.B ~Ermentrout}, {\em Strips or spots? Nonlinear effects in
%bifurcation of reaction diffusion equations on squrar}, Proc. R.
%Soc. Lond. A, 434, 413--417, 1991.

\bibitem{eck}
\textsc{Eckhaus, W.}
\textit{The Ginzburg--Landau manifold is an attractor}
J.~Nonlinear Science~3, 329--348, 1993.

\bibitem{g-m}
\textsc{Gierer, A. and H.~Meinhardt},
\textit{A Theory of Biologocal Pattern Formation},
Kybernetik~12, 30--39, 1972.

\bibitem{h-o}
\textsc{Haken, H. and H.~Olbricht},
\textit{Analytical Treatment of Pattern Formation
in the Gierer--Meinhardt Model of Morphogenesis},
J.~Math.\ Biol.~6(4), 1978.

\bibitem{h}
\textsc{Henry, D.},
\textsl{Geometric theory of semilinear parabolic equations},
Lecture Notes in Mathematics, Vol.~840,
Springer-Verlag, Berlin, 1981.

\bibitem{hoyle}
\textsc{Hoyle, R.}
\textsl{Pattern Formation, An Introduction to Methods},
Cambridge University Press, 2006.

\bibitem{iww}
\textsc{Iron, D., J.~Wei, and M.~Winter},
\textit{Stability analysis of Turing patterns generated by the
Schnakenberg model},
J.~Math.\ Biol.~49 (4), 358--390, 2004.

\bibitem{k-m}
\textsc{Koch, A.J. and H.~Meinhardt},
\textit{Biological Pattern Formation},
Rev.\ Mod.\ Phys.~66, 1481--1508, 1994.

\bibitem{mw}
\textsc{Ma, T. and S.~Wang},
\textit{Dynamic Bifurcation of Nonlinear Evolution Equations},
Chin.\ Ann.\ Mathematics~26, 185--206, 2005.

\bibitem{mw-b}
\textsc{Ma, T. and S.~Wang},
\textsl{Bifurcation Theory and Applications},
World Scientific, 2005.

\bibitem{mpf}
\textsc{Meinhardt, H., P.~Prusinkiewicz, and D.~R.~Fowler},
\textsl{The Algorithmic Beauty of Sea Shells},
Springer-Verlag, 2003.

\bibitem{murray}
\textsc{Murray, J.~D.},
\textsl{Mathematical Biology II}, third ed.,
Springer-Verlag, 2003.

%\bibitem{nm}
%{\sc B.N.~Nagorcka, J.R.~Mooney}, {\em From strips to spots:
%prepatterns which can be produced in the skin by reaction-diffusion
%systems}, IMA J. Math. Appl. Med. and Biol.~9, 249--267, 1992.

\bibitem{ni-1}
\textsc{Ni, Wei-Ming},
\textit{Diffusion, cross-diffusion, and their spike-layer steady states},
Notices Am.\ Math.\ Soc.~45, 9--18, 1998.

\bibitem{ni-2}
\textsc{Ni, Wei-Ming, S.~Kanako, and I.~Takagi},
\textit{The dynamics of a kinetic activator--inhibitor system},
J.~Diff.\ Eq.~229, 426--465, 2006.

\bibitem{pmm}
\textsc{Page, K.~M., P.~K.~Maini, and N.~A.~M.~Monk},
\textit{Pattern formation in spatially heterogeneous
Turing reaction--diffusion models},
Physica D~181, 80--101, 2002.

\bibitem{pmm-2005}
\textsc{Page, K.~M., P.~K.~Maini, and N.~A.~M.~Monk},
\textit{Complex pattern formation in reaction--diffusion
systems with spatially vaying diffusion coefficients},
Physica D~202 95--115, 2005.

\bibitem{sch}
\textsc{Schnakenberg, J.},
\textit{Simple Chemical Reaction Systems with Limit Cycle Behavior},
J.\ Theor.\ Biol.~81, 389--400, 1979.

\bibitem{schn-1}
\textsc{Schneider, G.},
\textit{Global existence via Ginzburg--Landau formalism
and pseudo-orbits of G-L approximations},
Comm.\ Math.\ Phys.~164, 157--179, 1994.

\bibitem{schn-2}
\textsc{Schneider, G.},
\textit{Nonlinear diffusive stability of spatially-periodic
solutions--abstract theorem and higher space dimensions},
Tohoku Math.~J.~8, 159--167, 1998.

\bibitem{smoller}
\textsc{Smoller, J.},
\textsl{Shock Waves and Reaction--Diffusion Equations}, second ed.,
Springer-Verlag, 1994.

\bibitem{t}
\textsc{Takagi, I.},
\textit{A priori estimates for stationary solutions
of an activator--inhibitor model due to Gierer and Mainhardt},
Tohoku Math.~J.~(2), 34(1), 113--132, 1982.

\bibitem{tu}
\textsc{Turing, A.},
\textit{The Chemical Basis of Morphogenesis},
Philos.\ Trans.\ Roy.\ Soc.\ London B~237, 37--52, 1952.

\bibitem{vh}
\textsc{van Harten, A.}
\textit{On the validity of the Ginzburg--Landau equation}
J.~Nonlinear Science~1, 397--422, 1991.

\bibitem{w}
\textsc{Wollkind, D.~J., V.~S.~Manoranjan, and L.~Zhang},
\textit{Weakly Nonlinear Stability Analyses of Prototype
Reaction--Diffusion Model Equations},
SIAM Rev.~36, 176--214, 1994.

\bibitem{z-m}
\textsc{Zhu, M. and J.D.~Murray},
\textit{Parameter Domains for Generating Spatial Pattern},
Int.\ J.\ Bifurcation and Chaos~5, 1503--1524, 1995.

%\bibitem{t}
%{\sc Temam, R.},
%{\em Infinte-Dymensional Dynamical systems in Mechanics and Physics},
%Second Edition, Springer (1997).

\end{thebibliography}
\end{document}